\documentclass{aa}
\usepackage{graphicx} 
\usepackage[varg]{txfonts}
\usepackage{natbib,twoopt}
\usepackage{adjustbox}
\usepackage{soul}
\usepackage{float}
\usepackage{placeins}
\usepackage{longtable} 
\usepackage{lscape}    
\usepackage[breaklinks=true]{hyperref} 
\bibpunct{(}{)}{;}{a}{}{,}             
\makeatletter
  \newcommandtwoopt{\citeads}[3][][]{\href{http://adsabs.harvard.edu/abs/#3}%
    {\def\hyper@linkstart##1##2{}%
     \let\hyper@linkend\@empty\citealp[#1][#2]{#3}}}
  \newcommandtwoopt{\citepads}[3][][]{\href{http://adsabs.harvard.edu/abs/#3}%
    {\def\hyper@linkstart##1##2{}%
     \let\hyper@linkend\@empty\citep[#1][#2]{#3}}}
  \newcommandtwoopt{\citetads}[3][][]{\href{http://adsabs.harvard.edu/abs/#3}%
    {\def\hyper@linkstart##1##2{}%
     \let\hyper@linkend\@empty\citet[#1][#2]{#3}}}
  \newcommandtwoopt{\citeyearads}[3][][]%
    {\href{http://adsabs.harvard.edu/abs/#3}
    {\def\hyper@linkstart##1##2{}%
     \let\hyper@linkend\@empty\citeyear[#1][#2]{#3}}}
\makeatother

\title{A multi-frequency{\bf ,} multi-epoch radio continuum study of the Arches cluster with the {\bf VLA}}
\author{
        M. Cano-Gonz\'alez
         \inst{1}
          \and
        R. Sch\"odel
         \inst{1}
          \and
          A. Alberdi
         \inst{1}
         \and
         J. Moldón
         \inst{1}
          \and
          M. P\'erez-Torres
         \inst{1,2}
         \and
          F. Najarro
         \inst{3}
         \and
         A.T. Gallego-Calvente
         \inst{1,4}
}
\institute{
           Instituto de Astrof\'isica de Andaluc\'ia (CSIC),
           Glorieta de la Astronom\'ia s/n, 18008 Granada, Spain.
           \email{mcano@iaa.es}
           \and
           School of Sciences, European University Cyprus, Diogenes street, Engomi, 1516 Nicosia, Cyprus
           \and
           Centro de Astrobiolog\'ia, CSIC-INTA, Ctra de Torrej\'on a Ajalvir km 4, 28850 Torrej\'on de Ardoz, Madrid, Spain
           \and 
           Universitat de València. Departament d'Astronomia i Astrofísica. Facultat de Física. Av. Doctor Moliner, 50 - Burjassot, Valencia 
}
\date{}

\begin{document}
\abstract
   {The Arches cluster, one of the most massive clusters in the Milky Way, is located about 30 pc in projection from the central massive black hole Sagittarius A$^*$. With its high mass, young age, and location in the Galaxy’s most extreme star forming environment, the Arches is an extraordinary laboratory to study massive stars and clusters.}
   {Our objective is to improve our knowledge of the properties of massive stars and 
  the Arches cluster through high angular resolution radio continuum studies.}
   {We observed the Arches cluster with the Karl G. Jansky Very Large Array in the C- and X-bands (central frequencies of 6 and 10 GHz respectively) in five epochs at X-band and two epochs at C-band observations throughout 2016, 2018, and 2022, covering time spans ranging from 22 days to six years. We used the A-configuration to achieve the highest possible angular resolution and cross-matched the detected point sources with stars detected in the infrared, using proper motion catalogues to ensure cluster membership. }
   {We report the most extensive radio point source catalogue of the cluster to date, with a total of 25 radio detections (seven more than the most recent study). We also created the deepest ($2.5\, \mu\mathrm{Jy}$ in X-band) images of the cluster so far in the 4 to 12 GHz frequency range. Most of our stellar radio sources (12/18) show a positive spectral index, indicating that the dominant emission process is free-free thermal radiation, which probably originates from stellar winds. We found that radio variability is more frequent than what was inferred from previous observations, affecting up to $60\%$ of the sources associated with bright stellar counterparts, with two of them, F18 and F26 showing extreme flux variability. We propose four of our detections (F6, F18, F19 and F26) as primary candidates for colliding-wind binaries based on their consistent flat-to-negative spectral index. We classify F7, F9, F12, F14, and F55 as secondary colliding wind binary candidates based on their high flux and/or spectral index variability, and X-ray counterparts. Thus, we infer a $14/23\approx61\%$ multiplicity fraction for the Arches cluster radio-stars when combining our findings with recent infrared radial velocity studies.}
   {}

\maketitle

\section{Introduction}

Young massive star clusters (YMCs) are targets of great interest in astrophysics. They can host from dozens to well over a hundred massive $(\gtrsim10 M_\sun)$ early-type stars which dramatically affect their environment with their intense UV radiation, strong stellar winds, and dramatic demise as supernovae. Since massive stars are short-lived, there are only a handful of  YMCs with masses above $\sim10^4\, M_\odot$ known in the Milky Way, all of them located at several kpc distance from Earth. As concerns extragalactic YMCs, beyond the Milky Way and Magellanic Clouds, their individual members cannot be easily separated and are much fainter than their Galactic counterparts.
The Galactic Centre (GC, at a distance of $\approx8$ kpc, \citealp{GRAVITY2019}) is host to three YMCs: the Arches \citep{Figer2002} and Quintuplet clusters \citep{Figer1999}, and the $10^4\, M_\odot$ of young massive stars within $0.5$ pc of the supermassive black hole \citep{Von_Fellenberg2022,Lu2013}. Arches and Quintuplet are located around 30 pc in projection from Sgr A*, within the so-called central molecular zone (CMZ), the turbulent, inner $R\sim100$ pc of the Milky Way \citep{Henshaw2023}. Both clusters are around $2-5$ Myr old, with masses of a few $10^4\,M_\sun$ \citep{Najarro2004,Martins2008,Liermann2012,Clark_I,Clark2018_Quintuplet}.

The CMZ contains $3-10\%$ of the molecular gas in the Milky Way, despite occupying $\leq0.01\%$ of its volume. The GC is the most prolific star forming region in the entire Galaxy and may be considered a local analogue of high-redshift starburst galaxies \citep{Kruijssen2013}. Our position within the Galactic disc makes the GC an extremely extinguished target ($A_V\sim30$ mag, \citealp[e.g.~][]{Nishiyama2008,GALACTICNUCLEUS_III}). Thus, studies of the Arches and Quintuplet clusters are limited to the radio, infrared, and X-ray frequency domains \citep[e.g.~][]{G-C2021,Clark_I,Muno2009}.

Radio emission from massive stars is associated with their strong ionised winds. There are two main non-exclusive emission mechanisms that cause radio emission from massive stars in the few GHz range. On the one hand, thermal free-free radiation is emitted by the so-called line-driven ionised winds of massive stars \citep{castor1975, Abbott1980}. On the other hand, in a system of massive binaries, non-thermal synchrotron emission may arise from the so-called colliding wind region between the stars. In order for this non-thermal emission to dominate over the free-free radiation arising from the winds of each star, the orbital separation must be sufficiently large (periods of more than a few weeks), otherwise, the synchrotron emission will be absorbed within the optically thick area of the colliding wind region, thus masking the observational traces of binarity \citep{DeBecker2007, Sanchez-Bermudez2019}. Therefore, only a subset of all colliding wind binaries (CWBs) present observable traces of a particle-accelerating mechanism, which is considered to be the main mechanism driving non-thermal emission in massive stars \citep{DeBecker2007, DeBecker2013}.

With multi-frequency radio observations, we can measure the spectral index of a particular source, hence providing insight into its physical nature. Assuming a dependence of flux density with frequency of the form: $S_\nu\propto\nu^\alpha$, where $\alpha$ is the spectral index, we can discriminate between thermal (free-free) and non-thermal (synchrotron) emission. The former is characterised by a positive spectral index (assuming a spherically symmetric, homogeneous wind, arising from a single star: $\alpha\approx0.6$ \citealp{Wright_Barlow1975}). The latter is associated with a significantly lower spectral index ($\alpha\lesssim0.0$, \citealp{DeBecker2007}). Furthermore, multi-epoch observations can provide a hint into the orbital phase of a multiple system by studying changes in the spectral index of a particular radio-star \citep{Dougherty2005, Sanchez-Bermudez2019}. Highly eccentric systems, which the Arches is suggested to host \citep{Clark_IV}, can be specially prone to spectral index variability.

Deviations from the $\alpha\approx0.6$ canonical value for free-free thermal emission are expected and can be caused by wind clumping and flares, a thermal contribution from a short-period CWB, or due to a contribution from non-thermal radiation within the colliding wind region of a binary system. However, thermal radio emission arising from stable, fully ionised, isolated stars is not expected to display relevant flux variability on timescales shorter than the star's evolutionary timescale, with the exception of the luminous blue variable phase. Therefore, radio variability may be an indirect indicator of a non-negligible, non-thermal component in the observed radio emission \citep{DeBecker2013} or (less likely in the case of Wolf-Rayet stars) of an eruptive mass-loss episode.

Massive stars are scarce and located at relatively large (kpc) distances. Thus, massive stellar evolution, specially the post main-sequence phase, is an area of ongoing development, both theoretical and observational \citep[e.g.~][]{Andrews2019,Bjorklund2023,Toala2024}. We can use radio observations to detect free-free emission arising from stellar winds of massive stars, that is sensitive to mass-loss, which plays a crucial role in their evolutionary pathway, providing hints into wind variability, geometry (clumping) and flaring events. As previously mentioned, evidence of non-thermal radio emission is a well established indicator of multiplicity \citep{DeBecker2013}, making radio continuum observations a complementary approach to spectroscopic studies to study the binary fraction of massive stars. Moreover, whereas spectroscopic studies are more sensitive to short-period binaries (periods of a few days), non-thermal emission traces multiplicity in regions of the orbital parameter space possibly not available to radial velocity studies, given the limited length of the spectroscopic campaigns themselves. In addition, non-thermal emission is not affected by the orbital inclination of the system with respect to the line of sight, which is an important bias to take into account in spectroscopic studies.

In our previous work \citep{G-C2021}, we studied the radio emission of the Arches cluster stellar members with the aim of establishing constraints on the properties of cluster members with radio observations only. With a relatively limited time span (only two observational epochs to our disposal: 2016 and 2018), and most of the observations carried out at one band (X-band), our variability, spectral indices and mass-loss rates analysis left room for improvement. In this work, we build on such results by incorporating new multi-wavelength observations and proper-motions measurements, quantitatively improving our variability, mass-loss and spectral index analysis. This paper is structured as follows: Section \ref{section_obs} describes the data reduction and imaging process, Sect. \ref{section_analysis} shows how we extracted and analysed flux densities, source positions, spectral indices as well as the main results of the study. Finally, Sect. \ref{sect_discussion} elaborates on our results and Sect. \ref{sect_conclusions} sums up our conclusions.

\section{Observations and imaging}\label{section_obs}
We observed the Arches cluster (pointing position of $\alpha_{\mathrm{J2000}}=$17$^{\mathrm{h}}$45$^{\mathrm{m}}$$50.49^{\mathrm{s}}$, $\delta_{\mathrm{J2000}} = -28^{\degr}49\arcmin19.92\arcsec$) with the Karl G. Jansky Very Large Array (VLA) in A-configuration in the C- and X-bands (central frequencies of $6$ and $10$ GHz respectively, both with a total bandwidth of 4 GHz and 3-bit sampling). Our seven observations were carried out in three different years: 2016, 2018 and 2022, see Table \ref{table_observations} for details.

We used CASA 6.4.1 (Common Astronomy Software Applications, \citealp[]{CASA2022}) with the pipeline version 2022.2.0.64, and performed a standard continuum calibration of the data. We also calibrated our data manually for comparison, but the results in terms of point source flux extraction and image quality were extremely similar. The data from 2016 and 2018 are identical to those described in \citet{G-C2021}, but we re-reduced and calibrated them from scratch (see Sect. \ref{subsect_pilotstudy}). We used 3C286 (J1331+3030) as flux and bandpass calibrator for all observations. For the  2016 and 2018 observations we used  J1744--3116 as phase calibrator, whereas we used J1820--2528 for 2022. Such change was done because J1744--3116 showed some extended emission in A-configuration, but it did not result in any systematic flux difference between 2016, 2018 and 2022. Since our main goal is to study radio point sources, we reduced the contribution of the extended emission present in our field \citep[see e.g.~][]{Lang2005, Heywood_2022} by performing a \textit{u-v} cut of $100\,\mathrm{k}\lambda$ and $150\,\mathrm{k}\lambda$ for the C- and X-bands, respectively. This way we could severely diminish the contribution of the extended emission present in the radio-bright zone of the GC. Although said contribution was not fully removed, and some areas near the cluster core are still affected by it. This mainly results in higher rms noise in this particular region, specially in C-band.

We performed phase-only self-calibration for each individual observation as well as for the combined data. We combined polarisations and spectral windows in order to achieve sufficient signal-to-noise ratio (S/N) in the self-calibration solutions. The optimal solution intervals ($\la5\%$ solutions with $\mathrm{S/N}<5$) ranged from 20\,s to 30\,s for all our observations. For the concatenated data of 2016 and 2022, we used the largest solution interval of the individual observations composing them. We used the aforementioned \textit{u-v} cut to compute the self-calibration solutions with the \texttt{gaincal} task and to perform the subsequent imaging. We note that amplitude self-calibration  did not improve the image quality or flux extraction in any significant way. In fact, enforcing  amplitude self-calibration incremented the total amount of flagged visibilities, which resulted in a poorer image quality, especially around the phase centre, so we discarded that approach.

We deconvolved the images with the CLEAN algorithm \citep{Hogbom74}, implemented in the CASA task \texttt{tclean}. We used 4 pixels across the synthesised beam FWHM, which resulted in a pixel size of $0\farcs0825$ and $0\farcs05$ for the C- and X-band images respectively. Although the VLA primary beam's FWHM (in arcminutes) is $\approx 42\times(1\mathrm{GHz}/\nu)$, for this work, we mainly focused on the $1\arcmin\times1\arcmin$ area centred on the Arches cluster, so that we covered the entire massive star catalogue of \citet{Clark_I}. We tried different weightings: natural, uniform and Briggs with several robust parameters. We obtained the best results with Briggs weighting and a robust parameter of 0.5 in terms of point source sensitivity, resolution, and rms noise. The gain parameter was set to 0.05 as in \citet{G-C2021} to facilitate spotting and masking of the faintest sources during the interactive cleaning. We imposed an rms-based stopping criterion of $5\sigma$ for all images. Table\,\ref{table_observations} provides a list of the beam geometry and rms noise reached in all images.

\begin{figure*}
   \sidecaption
   \includegraphics[width=12cm]{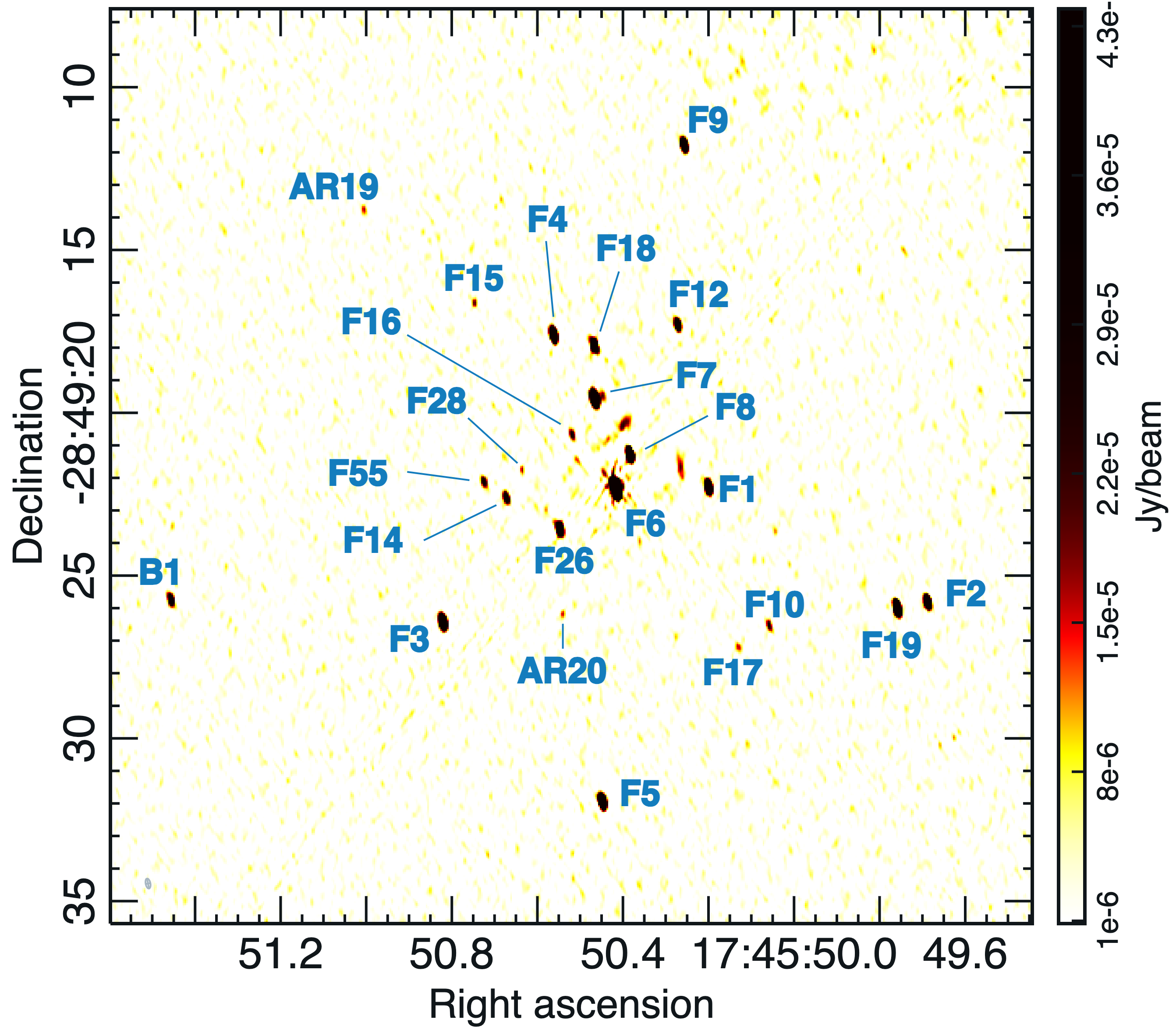}
      \caption{Deep X-band image of the Arches cluster (see Table \ref{table_observations} for details). Sources are labelled with the NIR stellar ID from \citet{Clark_I} with the exception of the AR19 and AR20 sources (see Sect. \ref{subsection_ID}).}
         \label{Arches_deep_X}
   \end{figure*}

Furthermore, we obtained deep radio images for each band by combining all visibilities of a given band using the CASA task \texttt{concat}. After concatenation, we performed global phase-only self-calibration in order to align all sources. To this end, we used one solution interval per scan. 
Figure \ref{Arches_deep_X} shows the deep, X-band image of the cluster, along the detected sources (see Sect. \ref{subsection_ID}).


\section{Analysis and results}\label{section_analysis}

\subsection{Flux extraction and astrometry}
We used the CASA task \texttt{imfit} to extract point source fluxes, positions, and their related uncertainties \citep{Condon1997}. In short, this task fits a single 2D Gaussian to the emission lying within a given region, provided by the user as input. We used circular regions with $1\farcs5$ diameter so that a sufficiently large area was covered for local rms computation during the fit. We made sure the resulting fits were similar in shape to the synthesised beam geometry of each image. In some particularly faint cases (such as F55, see Sect. \ref{subsection_ID}) that did not match the beam shape (specially for some sources near the cluster core at C-band, which is more prominently affected by the non-thermal extended emission) the peak flux was used instead of the integrated. Increasing the region size in \texttt{imfit} did not result in a significant difference in fluxes and their uncertainties. In addition to the flux error returned by the \texttt{imfit} task, we quadratically added systematic flux errors of $3\%$ and $5\%$ due to absolute calibration for X- and C-band respectively \citep{Perley_Butler2017}.

In a radio-interferometer, astrometric information is encoded in the phases. Since the data were self-calibrated in phases, absolute astrometric information was lost and we found systematic coordinate offsets between all our observations. We referred all individual images to our deep X-band image, given that the position uncertainties returned by the \texttt{imfit} task depend in part on the S/N of a given detection, and the deep X-band image shows the best off-source rms noise of all our images. Once the coordinates of each individual observation were referred to our deep X-band image, we used the Arches stellar catalogue provided by \citet{Hosek2022} to establish the astrometric reference frame. After correcting for an initial coordinate offset, our sources are cross-matched with their cluster members ($P_\mathrm{clust}>0.7$) within a median $0\farcs0003\pm0\farcs0039$ in R.A. and $-0\farcs001\pm0\farcs013$ in declination. The uncertainties represent the standard deviation of the previous offsets. Once we cross-matched our sources, all coordinate uncertainties were combined with the following expression:
\begin{equation}
    \sigma_{\mathrm{pos}} = \sqrt{\sigma_{\mathrm{fit}}^2 + \sigma_{\mathrm{std}}^2 + \sigma_{\mathrm{\theta}}^2}
\end{equation}
where $\sigma_{\mathrm{fit}}$ is the positional uncertainty returned by the fit in our deep X-band image, $\sigma_{\mathrm{std}}$ is the previously mentioned standard deviation of the offsets with respect to \citet{Hosek2022}, and $\sigma_{\mathrm{\theta}}=r_\theta\times\frac{\nu_c}{\nu_0}$ is a factor that takes into account the channel width ($\nu_c=2000\,\mathrm{kHz}$), central frequency $(\nu_0=10\,\mathrm{GHz})$, and $r_\theta$, which is the distance of a given source from the phase centre \citep{Thompson2017}. 

We present the most complete radio point source catalogue and the deepest radio images (in the few GHz range) of the Arches cluster to date. Flux densities, positions, and their related uncertainties can be found in Table \ref{fluxtable}. 

\subsection{Cross-matching with stars}\label{subsection_ID}
We found a systematic coordinate offset of $\approx1\arcsec$ between our deep X-band detections and the \citet{Clark_I} stellar catalogue. We computed median offsets of $-0\farcs92\pm0\farcs03$ in R.A. and $0\farcs483\pm0\farcs008$ in declination between our bright ($>0.1$ mJy), isolated deep X-band detections and the near-infrared (NIR) sources from \citet{Clark_I}. The errors represent the standard deviation of the offsets. After correcting for the offsets, radio-NIR cross-matching was performed by using a search radius of $0\farcs2$ around all our radio sources. Additionally, we used the HST Paschen-$\alpha$ catalogue from \citet{Dong2011} for the case of two radio sources with clear stellar counterparts that are not included in \citet{Clark_I}. In all, 23 out of the 25 point sources in the deep X-band image have bright stellar counterparts. The first column of Table \ref{fluxtable} shows the NIR stellar identification of our radio point sources, and Fig. \ref{Arches_deep_X} labels these sources in the deep X-band image.

\subsection{Variability assessment}\label{subsection_variability}
For a given detection and each band, we followed a similar procedure than the one described in \citet[see their appendix A1]{Zhao2020} to quantify radio-variability. We identified a particular source as variable if it met the following criteria:
\begin{equation}\label{eq_sigmaVAR}
    \sigma_{\mathrm{VAR}} = \frac{S_\nu^{\mathrm{max}} - S_\nu^{\mathrm{min}}}{\left( \sum_{i}^N\frac{1}{\sigma _{S_i}^2} \right)^{-1/2}}>5
\end{equation}
where $S_\nu^{\mathrm{max}}$ and $S_\nu^{\mathrm{min}}$ are, respectively, the maximum and minimum flux density values of a particular source at a given frequency $\nu$, $N$ is the number of observations in which a given source is detected, and $\sigma_{S_i}$ is the flux density uncertainty of said source in the $i$-eth observation. Therefore, we computed $\sigma_{\mathrm{VAR}}$ using five and two individual X- and C-band observations respectively. We note that, since the absolute systematic error due to absolute calibration affects all sources equally for a given band, we excluded it from the variability analysis. Using individual observations allows us only to measure the brightest sources. In order to obtain variability data of the faintest radio detections, we performed an analogous assessment with the combined 2016 and 2022 X-band images. Thus, we computed $\sigma_{\mathrm{VAR}}$ for 23 radio-stars, out of which 13 are noted as variables with this approach.  

\subsection{Spectral indices}\label{subsection_alphas}
We derived the spectral indices of our radio-stars using two different approaches.

\begin{figure}
  \resizebox{\hsize}{!}{\includegraphics{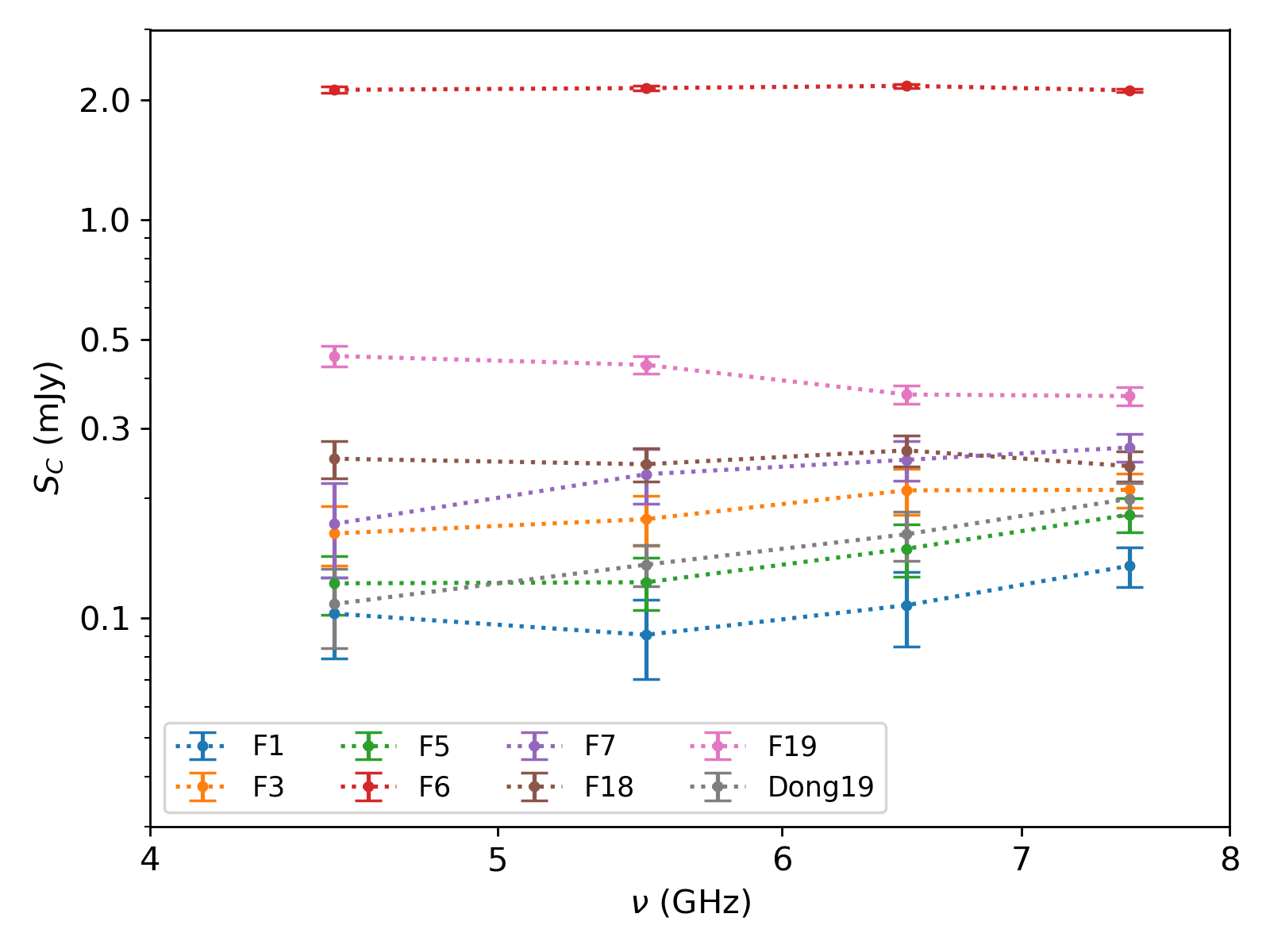}}
  \caption{Flux versus frequency plot in logarithmic space in the case of the C-band observations from 10 June 2018.}
  \label{fig_logS_vs_lognu}
\end{figure}

On the one hand, we divided the visibilities of each observation into four bandwidth chunks. Thus, we obtained four sub-images of 1 GHz bandwidth each. After measuring the fluxes of the sources in the sub-images, the spectral index of each source was computed via a least squares linear fit with an inverse variance weighting in logarithmic space (see Fig. \ref{fig_logS_vs_lognu}). In order for a spectral index to be computed this way, a source needed to be detected in all 1 GHz bandwidth images (minimum of four points for the linear fit). Furthermore, in the case of the C- and X-band observations carried out during 4 July 2022, we combined both measurement sets with the task \texttt{concat} and proceeded analogously. This resulted in a dataset with a broader bandwidth, ranging from 4 to 12 GHz, and thus a total of eight images of 1 GHz each, as well as a total of eight points to be fitted with the weighted least squares method.

We took into account both the fitting error returned by \texttt{imfit} as well the previously mentioned error due to absolute calibration when setting the weights of the least squares fit.

On the other hand, for those sources that were too faint to be fitted in the sub-images, and that were detected in both bands for a given year (2018 or 2022), we derived their spectral index as follows:

\begin{equation}\label{eq_alpha}
    \alpha = \frac{\log(S_C/S_X)}{\log(\nu_C/\nu_X)}
\end{equation}
and their related uncertainties through conventional error propagation:

\begin{equation}\label{eq_dalpha}
    \sigma_\alpha = \frac{1}{\log(\nu_C/\nu_X)}\times \sqrt{\left(\frac{\sigma_{S_C}}{S_C}\right)^2 + \left(\frac{\sigma_{S_X}}{S_X}\right)^2}
\end{equation}
where $S_\nu$ and $\sigma_{S_\nu}$ are, respectively, the flux and flux error of a particular source for a given frequency $\nu$, and $\nu_C$ and $\nu_X$ are the central frequencies of the C- and X-bands (6 and 10 GHz, respectively). We note that, if we use the latter method, since the C- and X-band 2018 observations were carried out two months apart, we must consider the possibility of source variability in this time span. If that were the case, the spectral index derived from this pair of observations would not be reliable. To address this issue, we used the two pairs of X-band observations from 2016 and 2022, which were taken a few months apart, and performed the same variability analysis described in Sect. \ref{subsection_variability}. Thus, we found that two sources: F6 and F18, exhibit significant variability on a few months time scale. This makes their 2018 spectral indices derived with Eqs. \ref{eq_alpha} and \ref{eq_dalpha} unreliable. 

Table \ref{table_alphas} shows the spectral indices of our radio-stars. In this Table, the columns denoted as $\alpha_{\mathrm{[band]}}^{\mathrm{[DD/MM/YY]}}$ correspond to the spectral indices derived from the least squares method and those columns noted as $\alpha^{\mathrm{[year]}}$ correspond to the values obtained from the error propagation method using Eqs. \ref{eq_alpha} and \ref{eq_dalpha}.

\def\arraystretch{1.2}
\setlength{\tabcolsep}{4pt}
\begin{table*}
\caption{Spectral indices}
\label{table_alphas}
\centering
\begin{tabular}{c c c c c c c c c}
\hline \hline
\noalign{\vskip 0.5mm}
ID  & $\alpha_X^{\mathrm{04/10/16}}$ &  $\alpha_X^{\mathrm{26/10/16}}$ & $\alpha_X^{\mathrm{11/04/18}}$ & $\alpha_C^{\mathrm{10/06/18}}$ & $\alpha^{\mathrm{2018}}$ & $\alpha_X^{\mathrm{07/05/22}}$ & $\alpha_{CX}^{\mathrm{04/07/22}}$ & $\alpha^{\mathrm{2022}}$ \\
\noalign{\vskip 0.5mm}
\hline
F6 & $0.01 \pm 0.06$ & $-0.02 \pm 0.07$ & $-0.09 \pm 0.03$ & $0.01 \pm 0.04$ & -- & $-0.26 \pm 0.05$ & $-0.15 \pm 0.02$ & $-0.17 \pm 0.12$ \\
F7 & $0.43 \pm 0.04$ & $0.48 \pm 0.12$ & $0.82 \pm 0.08$ & $0.79 \pm 0.17$ & $0.79 \pm 0.24$ & $1.08 \pm 0.09$ & $0.90 \pm 0.04$ & $0.92 \pm 0.21$ \\
F19 & $-0.75 \pm 0.27$ & $-0.36 \pm 0.17$ & $-0.21 \pm 0.19$ & $-0.51 \pm 0.12$ & $-0.75 \pm 0.14$ & $-0.61 \pm 0.15$ & $-0.32 \pm 0.06$ & $-0.35 \pm 0.15$ \\
F3 & $0.46 \pm 0.17$ & $0.83 \pm 0.11$ & $0.74 \pm 0.17$ & $0.54 \pm 0.12$ & $0.56 \pm 0.17$ & $0.40 \pm 0.04$ & $0.85 \pm 0.10$ & $0.84 \pm 0.19$ \\
Dong19 & $0.56 \pm 0.42$ & $0.64 \pm 0.10$ & $0.88 \pm 0.10$ & $1.18 \pm 0.05$ & $0.68 \pm 0.18$ & $0.59 \pm 0.10$ & $0.80 \pm 0.14$ & $0.80 \pm 0.18$ \\
F4 & $0.91 \pm 0.21$ & $0.60 \pm 0.45$ & $1.24 \pm 0.13$ & -- & $1.02 \pm 0.18$ & $0.50 \pm 0.16$ & $0.66 \pm 0.15$ & $0.82 \pm 0.19$ \\
F5 & $0.66 \pm 0.06$ & $0.54 \pm 0.22$ & $0.40 \pm 0.08$ & $0.82 \pm 0.23$ & $0.60 \pm 0.17$ & $0.89 \pm 0.40$ & $0.87 \pm 0.12$ & $0.85 \pm 0.20$ \\
F8 & $0.25 \pm 0.14$ & $0.69 \pm 0.32$ & $0.96 \pm 0.32$ & -- & $0.44 \pm 0.41$ & $0.48 \pm 0.06$ & $1.10\pm0.08$ & $0.67 \pm 0.54$ \\
Dong96 & $0.98 \pm 0.74$ & $0.50 \pm 0.21$ & $0.87 \pm 0.15$ & -- & $0.73 \pm 0.21$ & $1.51 \pm 0.15$ & $0.76\pm0.14$ & $0.79 \pm 0.23$ \\
F1 & $0.58 \pm 0.46$ & $0.70 \pm 0.61$ & $-0.52 \pm 0.50$ & $0.62 \pm 0.35$ & $1.02 \pm 0.33$ & $0.71 \pm 0.22$ & $0.34 \pm 0.14$ & $0.93 \pm 0.37$ \\
F18 & $-0.45 \pm 0.32^{*}$ & -- & $0.04 \pm 0.29$ & $-0.02 \pm 0.13$ & -- & $-0.39 \pm 0.15$ & -- & $-0.32 \pm 0.49$ \\
F26 & $-1.01 \pm 0.14$ & $0.07 \pm 0.29$ & $-0.20 \pm 0.17$ & -- & -- & $-0.22 \pm 0.15$ & -- & -- \\
F2 & $0.34 \pm 0.49$ & $0.94 \pm 0.19$ & $0.71 \pm 0.76$ & -- & $0.58 \pm 0.23$ & $-0.07 \pm 0.32$ & $0.49 \pm 0.18$ & $0.45 \pm 0.34$ \\
F9 & $0.72 \pm 0.41$ & $0.95 \pm 0.26$ & $0.78 \pm 0.36^{*}$ & -- & $1.17 \pm 0.36$ & $0.63 \pm 0.34$ & $0.81 \pm 0.24$ & $1.03 \pm 0.42$ \\
F12 & $-0.27 \pm 0.51$ & -- & $0.22 \pm 0.38$ & -- & $0.59 \pm 0.37$ & -- & -- & $0.32 \pm 0.38$ \\
B1 & -- & -- & -- & -- & $0.92 \pm 0.49$ & -- & -- & $0.70 \pm 0.53$ \\
F14 & -- & -- & -- & -- & $-1.59 \pm 0.47$ & -- & -- & $0.70 \pm 0.62$ \\
F55 & -- & -- & -- & -- & $-1.15 \pm 0.90$ & -- & -- & $-0.86 \pm 0.81$ \\

\hline \hline
\end{tabular}
\tablefoot{\tablefoottext{*}{These values were obtained using the peak flux from each sub-image. Due to relatively high S/N, the Gaussian fits did not match the synthesised beam in some sub-images, which resulted in a flux overestimation while performing the least squares fit.}}
\end{table*}

In an similar way as with the flux densities, we searched for significant changes in $\alpha$ throughout our observations. To do that, we computed the following quantity:

\begin{equation}\label{eq_alpha_var}
    \Xi = \frac{ |\alpha_i - \alpha_j |}{\left(\frac{1}{\mathrm{e}\alpha_i^2} +  \frac{1}{\mathrm{e}\alpha_j^2}\right)^{-1/2}}
\end{equation}
where the sub-indices $i,j$ represent the two different observations in each year, and $\mathrm{e}\alpha_{i,j}$ denotes their respective spectral index uncertainties. We refer to Sect. \ref{sect_alphavar} for further discussion of the $\Xi$ values obtained for our radio detections.

\subsection{Proper-motions and cluster membership}
We used the proper motion catalogue by \citet{Hosek2022} (Arches absolute proper motions of $\mu_{\alpha}\,\cos{\delta} = -0.80\pm0.032\, \mathrm{mas\,yr}^{-1}$ and $\mu_\delta= -1.89\pm0.021 \,\mathrm{mas}\, \mathrm{yr}^{-1}$) to ensure cluster membership of our radio point-sources. Proper motions, as well as the cluster membership probability from their work are listed in Table \ref{table_pm} along the corresponding NIR stellar ID.

\def\arraystretch{1.2}
\begin{table}
\caption{Proper motions and cluster membership probability.} 
\label{table_pm}
\centering
\begin{tabular}{c c c c}
\hline \hline
ID  & $\mu_{\alpha}\,\cos{\delta}$ &  $\mu_\delta$ & $P_{\mathrm{cluster}}$ \\
\hline
 F6 & $-0.77\pm0.08$ & $-1.92\pm0.07$ & 0.94 \\
 F7 & $-0.49\pm0.08$ & $-2.08\pm0.06$ & 0.11 \\
F19 & $-0.93\pm0.07$ & $-1.78\pm0.06$ & 0.84 \\
 F3 & $-0.70\pm0.08$ & $-1.83\pm0.06$ & 0.92 \\
Dong19 & $-0.86\pm0.07$ & $-1.95\pm0.06$ & 0.94 \\
 F4 & $-0.70\pm0.08$ & $-1.90\pm0.06$ & 0.93 \\
 F5 & $-0.90\pm0.07$ & $-2.05\pm0.06$ & 0.78 \\
 F8 & $-0.93\pm0.08$ & $-1.67\pm0.06$ & 0.50 \\
 Dong96 & $-0.68\pm0.07$ & $-1.86\pm0.07$ & 0.91 \\
 F1 & $-1.04\pm0.10$ & $-2.01\pm0.09$ & 0.58 \\
F18 & $-0.83\pm0.08$ & $-1.97\pm0.07$ & 0.93 \\
F26 & $-0.81\pm0.08$ & $-1.87\pm0.06$ & 0.95 \\
 F2 & $-0.53\pm0.08$ & $-1.87\pm0.06$ & 0.54 \\
 F9 & $-0.82\pm0.07$ & $-1.97\pm0.06$ & 0.93 \\
F12 & $-0.70\pm0.11$ & $-1.85\pm0.10$ & 0.91 \\
 B1 & $-0.89\pm0.07$ & $-1.92\pm0.05$ & 0.93 \\
F14 & $-0.78\pm0.07$ & $-1.91\pm0.06$ & 0.95 \\
F55 & $-0.81\pm0.07$ & $-1.66\pm0.05$ & 0.60 \\
F16 & $-0.50\pm0.08$ & $-1.92\pm0.07$ & 0.39 \\
F10 & $-0.86\pm0.08$ & $-1.92\pm0.06$ & 0.94 \\
F15 & $-0.96\pm0.08$ & $-1.93\pm0.07$ & 0.86 \\
F17 & $-0.84\pm0.07$ & $-1.85\pm0.05$ & 0.95 \\
F28 & $-0.58\pm0.08$ & $-1.72\pm0.06$ & 0.46 \\
\hline \hline
\end{tabular}
\tablefoot{Values from \citet{Hosek2022}. Proper motion units in mas\, yr$^{-1}$}
\end{table}

\begin{figure}
  \resizebox{\hsize}{!}{\includegraphics{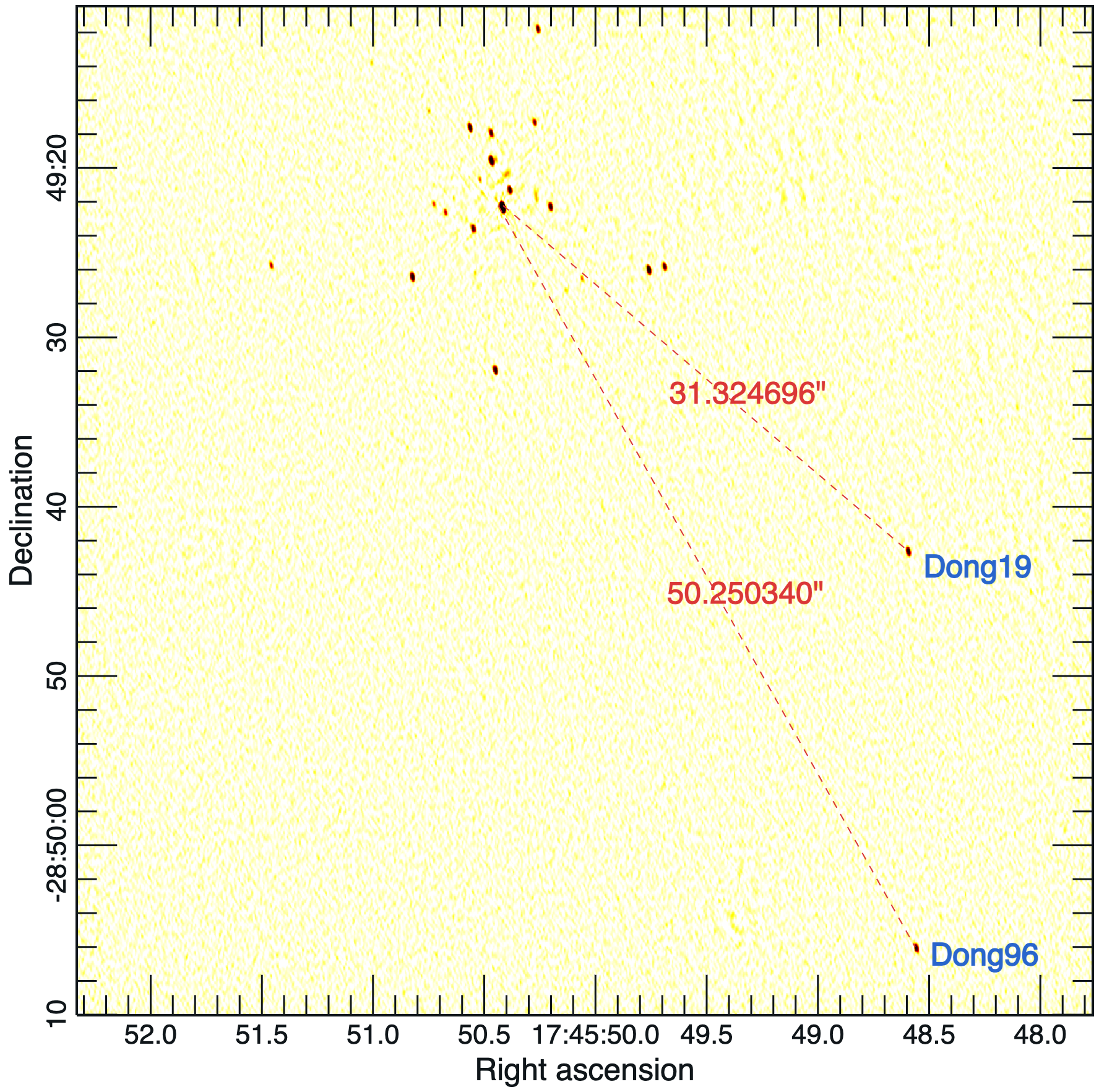}}
  \caption{Position of Dong19 and Dong96 with respect to the cluster core.}
  \label{fig_Dongs_distance}
\end{figure}

With these results, we could confirm that Dong19 and Dong96 are 'bona-fide' cluster members. Therefore, we included them in our analysis, despite being respectively separated $\approx30\arcsec$ and $\approx50\arcsec$ from the cluster core. Figure \ref{fig_Dongs_distance} shows the position of both sources with respect to the Arches. 

Table \ref{table_pm} also shows relatively low cluster membership probabilities for some of our detections, for example F7. We note that \citet{Hosek2022} cluster membership probability is based on a proper motion-only Bayesian analysis, but we also considered detections close to the Arches core as cluster members because the possibilities of random chance alignment of an isolated massive star with the cluster core are low \citep[e.g.~][]{Clark2021}.

\section{Discussion}\label{sect_discussion}
\subsection{Comparison with the pilot study}\label{subsect_pilotstudy} 
As explained in Sect. \ref{section_obs}, for the sake of consistency, we decided to reduce all data, including the 2016 and 2018 data, using the same version of the VLA pipeline. Our study presents two major improvements with respect to the pilot study of \citet{G-C2021}. First, while re-reducing the 2016 and 2018 data, we could achieve self-calibration, which drastically improved image quality. This resulted in more accurate flux extraction and better uncertainties, specially around the central, brightest source F6. Secondly, the addition of the 2022 data provided us with a longer temporal baseline to asses variability, as well as another measurement of the spectral index and the deepest X-band images of any year. Also, \citet{Hosek2022} proper motions study allowed us to confirm Dong19 and Dong96 as cluster members based on their Bayesian analysis. All this has led to the detection of 25 radio sources, seven more than in \citet{G-C2021}, as well as an expansion upon their variability analysis, in which they find that $\lesssim15\%$ of their sources are variable, strongly contrasting with the new $\lesssim60\%$ variability fraction. Also, we did not have to correct for the factor $\sim2$ offset between the 2016 and 2018 X-band observations as done by \citet{G-C2021}, because we detected no systematic offset between these years. Finally, the approach to spectral index computation discussed in Sect. \ref{subsection_alphas} has allowed us to study the spectral behaviour of our radio-stars at all our epochs, improving the spectral index uncertainties by a factor of 2-3, specially for the brightest sources, which is relevant to determine their variability, their possible correlation with flux density variability, and any potential connection with the phase of a multiple system.

\subsection{Spectral index variability}\label{sect_alphavar}
\begin{figure}
  \resizebox{\hsize}{!}{\includegraphics{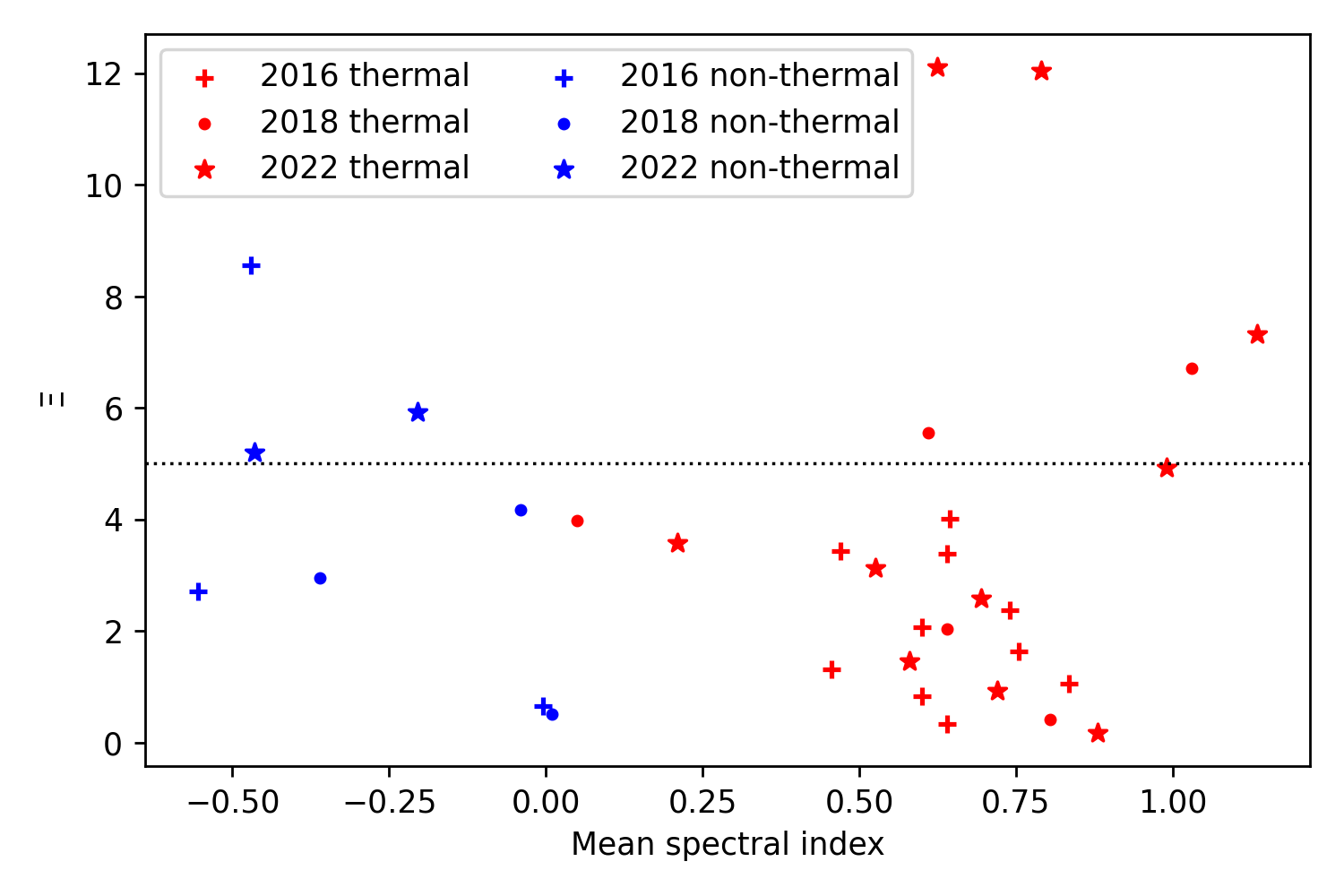}}
  \caption{Spectral index variation between observations in a given year as a function of mean spectral index. Red and blue markers represent thermal and non-thermal sources respectively. The horizontal dashed line represent the chosen variability threshold.}
  \label{fig_Xi_vs_alpha_epochs}
\end{figure}

Significant changes of the spectral index might be an indicator of changes in the relative weights of free-free and non-thermal emission, of an eruptive mass-loss episode or of the clumping and porosity in the stellar winds. Figure \ref{fig_Xi_vs_alpha_epochs} shows the $\Xi$ value from Eq. \ref{eq_alpha_var} as a function of mean spectral index. We classified sources as non-thermal if they showed negative or flat spectral indices consistently throughout our observations, and thermal sources as those with consistent positive $\alpha$ values throughout all epochs.

We set a threshold at $\Xi>5$, as we did with flux variability, and we can see from Fig. \ref{fig_Xi_vs_alpha_epochs} that, from a percentage standpoint, non-thermal sources are more likely to showcase spectral index variability between observations carried out in the same year (such is the case of F6, F26 and F19). This behaviour may be caused by the observed orbital phase of a binary or multiple system, during which the non-thermal synchrotron component becomes more or less prevalent \citep{Dougherty2005, DeBecker2007, Sanchez-Bermudez2019}. Also, we can see from Fig. \ref{fig_Xi_vs_alpha_epochs} that thermal emitters tend to cluster around $\overline{\alpha}\sim0.7$ (near the canonical $\alpha\sim0.6$ value from \citealt{Wright_Barlow1975}) and $\Xi\sim2$, suggesting that they are less likely to display spectral index variability in timescales of a few months.

However, we can also see that some thermal sources showcase significant spectral index variability (that is the case for F8 or Dong96 during the observations carried out during 2022). Given that their spectral indices do not show changes below the canonically thermal $\alpha\approx0.6$ value at any epoch, these changes are possibly not due to a non-thermal emission contribution, but a change in the clumping structure and over-densities in the radio formation region. 

In order to check whether or not four points are enough to properly characterise the spectral response of our sources, we divided the X-band 7 May dataset into eight 4-spectral window chunks (500 MHz each chunk) and performed the same least squares fit as in the sub-images from Sect. \ref{subsection_alphas}. This resulted in a total of eight points to be fitted by a polynomial degree one. We obtained similar $\alpha$ values for our sources, but with a factor of 2 higher uncertainties. Therefore, we decided to use the 8-spectral window chunks (four total data points) as a better method of characterising the spectral indices of the Arches radio-stars.

\subsection{Mass-loss rates and clumping ratios}
In principle, one can estimate the mass-loss rates of our thermal radio-stars using the prescriptions of \citet{Wright_Barlow1975}. Their model assumes a spherically symmetric wind, that only emits free-free emission. However, it is known that wind clumping can affect mass-loss rates estimates by factors of 3-4 \citep{Smith2014_review,Bjorklund2023}. Therefore, for our thermal radio-stars ($\alpha>0$), we computed the intrinsic mass-loss rates combined with the clumping factor ($f_\mathrm{cl}\equiv\langle \rho ^2 \rangle/\langle \rho \rangle^2$) with the following expression (as in e.g. \citealt{Andrews2019}):

\begin{equation}\label{eq_MLRs}
    \dot{M}\sqrt{f_\mathrm{cl}^\mathrm{[band]}} = 0.095\times\frac{\mu\, v_\infty\, S_\nu^{3/4}\, d^{3/2}}{Z\, (\gamma\, g_\nu\, \nu)^{1/2}}\, \, \,  \left[M_\sun\,\mathrm{yr^{-1}}\right]
\end{equation}
 with the free-free Gaunt factor $g_\nu$ defined by:
 \begin{equation}
     g_\nu = 9.77\left(1+0.13\log\left(  \frac{T_e^{3/2}}{Z\,\nu}\right)\right)
 \end{equation}
 where $\mu$ is the mean molecular weight per ion, $v_\infty$ is the terminal wind velocity in units of km s$^{-1}$ (both taken from \citealt{Martins2008}), $S_\nu$ is the flux density in Jy at the observing frequency $\nu$ in Hz, $d$ represents the distance to the target in kpc units (we assume a distance of 8 kpc to the GC), Z is the ratio of electron to ion density,  and $\gamma$ is the mean number of electrons per ion (both taken to be 1.0, as assumed in the Wolf-Rayet and O-type population of \citealt{Andrews2019}), finally, $T_e$, the electron temperature, is assumed to be $10^4$ K. Mass-loss rate uncertainties were calculated following the standard error propagation, ignoring the uncertainty related to the distance to the GC, for it systematically affects all cluster members in the same manner:

 \begin{equation}\label{eq_e_MLRs}
 \begin{split}
     \frac{\sigma_{\dot{M}}}{M_\sun\,\mathrm{yr}} = \frac{\dot{M}}{M_\sun\,\mathrm{yr}}\Bigg[\left(\frac{\sigma_{v_\infty}}{v_\infty}\right)^2 + \left(\frac{\sigma_\mu}{\mu}\right)^2 + \frac{9}{16}\left(\frac{\sigma_{S_\nu}}{S_\nu}\right)^2 + \\ 
     +  \left(\frac{\sigma_Z}{Z}\right)^2 + \frac{1}{4}\left(\frac{\sigma_\gamma}{\gamma}\right)^2 + \frac{1}{4}\left(\frac{\sigma_{g_\nu}}{g_\nu}\right)^2\Bigg]^{1/2}
 \end{split}
 \end{equation}

We assumed 0.08 dex uncertainties for $\gamma,Z,$ and $\mu$, and a 10\% relative error for the Gaunt factor as in \citet{G-C2021}.

We used the combined X-band data from 2016 and 2022, given that none of these sources showcased significant variability on timescales of months, and because the combined images provide smaller flux uncertainties. 

In addition, we also used the old VLA K-band data (central frequency of 22.5 GHz) from \citet{Lang2005} and computed the associated $\dot{M}\sqrt{f_{\mathrm{cl}}^K}$ values with eq. \ref{eq_MLRs}. K-band fluxes are useful because they trace radio emission closer to the stellar atmosphere, and because non-thermal emission is less prominent at higher frequencies \citep{Contreras1996}, allowing us to establish an upper limit for the $\dot{M}\sqrt{f_{\mathrm{cl}}^K}$ value of F6. Table \ref{table_MLRs} shows the $\dot{M}\sqrt{f_{\mathrm{cl}}^\mathrm{[band]}}$ factor derived for our thermal radio-stars in both bands and the three different years. 
 
\begin{table*}
\caption{$\dot{M}^{\mathrm{[YY]}}\sqrt{f_\mathrm{cl}^{\mathrm{[band]}}}$ factors for our sources.}
\label{table_MLRs}
\centering
\begin{tabular}{c c c c c c c}
\hline \hline
ID & $\dot{M}^{16}\sqrt{f_\mathrm{cl}^{X}}$ & $\dot{M}^{18}\sqrt{f_\mathrm{cl}^{X}}$ &  $\dot{M}^{18}\sqrt{f_\mathrm{cl}^{C}}$ & $\dot{M}^{22}\sqrt{f_\mathrm{cl}^{X}}$ &  $\dot{M}^{22}\sqrt{f_\mathrm{cl}^{C}}$ & $\dot{M}^{05}\sqrt{f_\mathrm{cl}^{K}}$ \tablefootmark{(a)} \\
\hline
    F6 & -- & -- & -- & -- & -- & $11.1 \pm 2.3$\\ 
    F7 & $5.6 \pm 0.9$ & $5.5 \pm 0.9$ & $5.1 \pm 0.9$ & $6.4 \pm 1.0$ & $5.4 \pm 0.9$ & $5.7 \pm 1.1$\\ 
    F3 & $3.4 \pm 0.5$ & $3.4 \pm 0.5$ & $3.4 \pm 0.5$ & $3.5 \pm 0.5$ & $3.2 \pm 0.5$ & $4.4 \pm 0.8$\\ 
Dong19\tablefootmark{(\textbf{b})} & $4.4 \pm 0.7$ & $4.3 \pm 0.7$ & $4.2 \pm 0.7$ & $4.6 \pm 0.7$ & $4.2 \pm 0.7$ & --\\ 
    F4 & $4.5 \pm 0.7$ & $4.6 \pm 0.7$ & $3.9 \pm 0.6$ & $5.1 \pm 0.8$ & $4.6 \pm 0.7$ & --\\ 
    F5 & $3.7 \pm 0.6$ & $3.6 \pm 0.5$ & $3.6 \pm 0.6$ & $3.8 \pm 0.6$ & $3.5 \pm 0.6$ & $4.7 \pm 1.0$\\ 
    F8 & $4.5 \pm 0.8$ & $4.2 \pm 0.7$ & $4.5 \pm 0.9$ & $4.4 \pm 0.7$ & $4.3 \pm 0.9$ & $6.7 \pm 1.2$\\ 
Dong96\tablefootmark{(\textbf{b})} & $3.7 \pm 0.6$ & $3.7 \pm 0.6$ & $3.5 \pm 0.6$ & $3.9 \pm 0.6$ & $3.5 \pm 0.6$ & --\\ 

    F1 & $2.6 \pm 0.4$ & $2.7 \pm 0.4$ & $2.3 \pm 0.4$ & $2.8 \pm 0.4$ & $2.4 \pm 0.5$ & --\\ 
    F2 & $3.3 \pm 0.5$ & $3.4 \pm 0.5$ & $3.4 \pm 0.6$ & $3.2 \pm 0.5$ & $3.2 \pm 0.6$ & --\\ 
    F9 & $4.4 \pm 0.7$ & $2.6 \pm 0.4$ & $2.1 \pm 0.4$ & $2.4 \pm 0.4$ & $1.8 \pm 0.4$ & --\\ 
    B1 & $1.4 \pm 0.2$ & $1.6 \pm 0.3$ & $1.4 \pm 0.3$ & $1.5 \pm 0.3$ & $1.4 \pm 0.3$ & --\\ 
   F14 & $1.3 \pm 0.2$ & -- & -- & $0.9 \pm 0.3$ & $1.2 \pm 0.3$ & --\\ 
   F16 & $0.9 \pm 0.2$ & $0.8 \pm 0.2$ & -- & $0.6 \pm 0.1$ & -- & --\\ 
   F10 & $0.8 \pm 0.2$ & $1.1 \pm 0.3$ & -- & $0.8 \pm 0.2$ & -- & --\\ 
   F15 & $0.9 \pm 0.2$ & -- & -- & $0.9 \pm 0.2$ & -- & --\\ 
   F28 & -- & -- & -- & $0.9 \pm 0.2$ & -- & --\\ 
\hline \hline
\end{tabular}
\tablefoot{All mass-loss rates and their uncertainties are in $10^{-5}\,M_\odot\,\mathrm{yr}^{-1}$ units.\tablefoottext{a}{We used the K-band old VLA data from \citet{Lang2005}.}
\tablefoottext{b}{For these sources we used $\mu=2$ and $v_\infty=1200\, \mathrm{km\,s}^{-1}$, which are typical for stars of their stellar type \citep[see, e.g.~][]{Andrews2019}.}}
\end{table*}

Once the clumped mass-loss rates are associated with our thermal sources, we can gain some basic insight into their wind geometry by computing the clumping ratios $f_{\mathrm{cl}}^{\nu_2}/f_{\mathrm{cl}}^{\nu_1}$, that follow from Eq. \ref{eq_MLRs}:

\begin{equation}\label{eq_clump_ratios}
     \frac{f_{\mathrm{cl}}^{\nu_2}}{f_{\mathrm{cl}}^{\nu_1}} = \frac{g_{\nu_1}}{g_{\nu_2}}\,  \frac{\nu_1}{\nu_2}  \left( \frac{S_{\nu_2}}{S_{\nu_1}}  \right)^{3/2}  \mathrm{where} \,\nu_1>\nu_2
 \end{equation}

 Again, we also used K-band fluxes from \citet{Lang2005} in Eq. \ref{eq_clump_ratios} to better sample the radio-photosphere at different distances from the stellar surface. Table \ref{table_fclump} shows the clumping ratios derived for our thermal sources.

\begin{table}
\caption{Clumping ratios}
\label{table_fclump}
\centering
\begin{tabular}{c c c c c c}
\hline \hline
ID  & $\left(f_\mathrm{cl}^C/f_\mathrm{cl}^X\right)_{18}$ & $\left(f_\mathrm{cl}^C/f_\mathrm{cl}^X\right)_{22}$\tablefootmark{a} &  $\left(f_\mathrm{cl}^C/f_\mathrm{cl}^K\right)$\tablefootmark{b} & $\left(f_\mathrm{cl}^X/f_\mathrm{cl}^K\right)$\tablefootmark{b} \\
\hline

     F7 & $0.9 \pm 0.2$ & $0.8 \pm 0.1$ & $0.9 \pm 0.3$ & $1.2 \pm 0.3$ \\ 
     F3 & $1.0 \pm 0.1$ & $0.8 \pm 0.1$ & $0.6 \pm 0.1$ & $0.6 \pm 0.2$ \\ 
 Dong19 & $0.9 \pm 0.1$ & $0.9 \pm 0.1$ & -- & -- \\ 
     F4 & $0.7 \pm 0.1$ & $0.8 \pm 0.1$ & -- & -- \\ 
     F5 & $1.0 \pm 0.1$ & $0.8 \pm 0.1$ & $0.6 \pm 0.2$ & $0.7 \pm 0.2$ \\ 
     F8 & $1.1 \pm 0.4$ & $0.9 \pm 0.4$ & $0.3 \pm 0.1$ & $0.4 \pm 0.1$ \\ 
 Dong96 & $0.9 \pm 0.1$ & $0.9 \pm 0.1$ & -- & -- \\ 
     F1 & $0.7 \pm 0.2$ & $0.8 \pm 0.2$ & -- & -- \\ 
     F2 & $1.0 \pm 0.2$ & $1.1 \pm 0.3$ & -- & -- \\ 
     F9 & $0.6 \pm 0.2$ & $0.7 \pm 0.2$ & -- & -- \\ 
     B1 & $0.8 \pm 0.3$ & $0.9 \pm 0.4$ & -- & -- \\ 
    F14 & -- & $0.9 \pm 0.4$ & -- & -- \\ 

\hline \hline
\end{tabular}
\tablefoot{
\tablefoottext{a}{2022 clumping ratios were computed using the contemporaneous set of C- and X-band data from 4 July.}\\
\tablefoottext{b}{We used the fluxes from the deep C- and X-band images to compute clumping ratios with \citet{Lang2005} K-band data.}
}

\end{table}

We can see from the first two columns of Table \ref{table_fclump} that, for most of our sources, clumping affects C- and X-bands similarly, while it may become more prevalent at K-band, which traces emission closer to the stellar surface. More multi-wavelength radio observations at higher frequencies (such as K-band) combined with infrared and sub-millimetre studies (carried out by instruments such as ALMA) may reveal a clearer picture on the radial dependence of clumping with observing frequency and thus help to establish stronger observational constraints on the intrinsic mass-loss of massive stars.

Although the clumping-scaled mass-loss rates broadly agree with existing literature \citep[e.g.~][Table 1]{Vink2022_review}, in this work we have decided to refrain from making a detailed comparison between our results and current stellar evolutionary models as it will be severely affected by uncertainties in extinction and clumping.

Differential extinction has been shown to be present throughout the GC \citep{schodel2010,GALACTICNUCLEUS_IV}. Recent studies in the Arches field \citep{Hosek2019} have suggested mean extinction values ($A_{Ks}\sim2.4$) which are lower  than those provided by \citet{Figer2002}, ($A_{Ks}\sim3.1$) and \citet{Martins2008}, ($A_{Ks}\sim2.8$). Furthermore, recent quantitative spectroscopy of individual stars confirms the presence of differential extinction in the Arches and identify large uncertainties in the resulting $A_{Ks}$ depending on the adopted extinction laws, leading up to 0.6 dex differences in the  stellar luminosities \citep[0.45 dex in mass-loss,][see their table 2]{Clark_IV}.

Furthermore, using the clumping values shown in \citet{Clark_IV} would be an oversimplification, because clumping is expected to be less prominent further away from the stellar surface (as shown in \citet{Runacres&Owoki2002} models and our clumping ratios). The structure and geometry of massive stellar wind clumps are topics of ongoing research \citep[e.g.~][]{Rubio-Diez2022} and a detailed observational analysis would require a multi-frequency, multi-facility study, which is left for future work and out of the scope of this paper.


\subsection{Nature of the sources}

\subsubsection{CWB candidates}

We may distinguish between primary and secondary CWB candidates. To fall under the former category, a given source must present a negative or flat spectral index consistently. However, significant flux variability could also be an indirect indicator of a particle accelerating mechanism at work, as expected in colliding-wind regions \citep[e.g.~][]{DeBecker2013}. Therefore, highly variable sources may be considered secondary candidates. Table \ref{table_varpar} shows different parameters that may be used to discriminate CWB among our radio detections.

\begin{table*}
\caption{Variability parameters}
\label{table_varpar}
\centering
\begin{tabular}{c c c c c c c}
\hline \hline
ID  & $\sigma_{\mathrm{VAR}}$\tablefootmark{(a)} &  $\overline{\alpha}\pm\sigma_\alpha$ & $\Xi_{2016}$\tablefootmark{(b)} & $\Xi_{2018}$\tablefootmark{(b)} & $\Xi_{2022}$\tablefootmark{(b)} & CWB candidate\tablefootmark{(c)} \\
\hline
\textbf{F6} & 56.3 & $-0.1 \pm 0.1$ & 0.7 & 4.2 & 5.9 & p \\
\textbf{F7} & 18.5 & $0.8 \pm 0.2$ & 1.3 & 0.4 & 4.9 & s \\
F19 & 28.6 & $-0.5 \pm 0.2$ & 2.7 & 3.0 & 5.2 & p\\
F3 & 6.6 & $0.6 \pm 0.2$ & 4.0 & 2.0 & 12.1 & No\\
Dong19 & 6.1 & $0.8 \pm 0.2$ & 0.8 & 6.7 & 2.6 & No \\
F4 & 15.0 & $0.8 \pm 0.3$ & 1.6 & -- & 1.5 & No \\
F5 & 5.5 & $0.7 \pm 0.2$ & 2.1 & 5.6 & 0.2 & No \\
\textbf{F8} & 1.0 & $0.7 \pm 0.3$ & 3.4 & -- & 12.1 & No \\
Dong96 & 4.4 & $0.9 \pm 0.3$ & 2.4 & -- & 7.3 & No\\
F1 & 4.5 & $0.4 \pm 0.4$ & 0.3 & 4.0 & 3.1 & No \\
\textbf{F18} & 49.0 & $-0.2 \pm 0.2$ & -- & 0.5 & -- & p \\
F26 & 10.1 & $-0.3 \pm 0.4$ & 8.6 & -- & -- & p \\
\textbf{F2} & 4.8 & $0.5 \pm 0.3$ & 3.4 & -- & 3.6 & No \\
F9 & 38.0 & $0.8 \pm 0.1$ & 1.0 & -- & 0.9 & s\\
F12 & 6.8 & $-0.0 \pm 0.2$ & -- & -- & --  & s \\
B1 & 4.8 & $0.8 \pm 0.1$ & -- & -- & -- & No \\
\textbf{F14} & 2.7 & $-0.5 \pm 1.2$ & -- & -- & -- & s\\
F55 & 4.7 & $-1.0 \pm 0.1$ & -- & -- & --  & s\\
\hline \hline
\end{tabular}
\tablefoot{Boldface in the ID indicates radial velocity binary candidate according to \citet{Clark_IV}. \tablefoottext{a}{Maximum value obtained from Eq. \ref{eq_sigmaVAR} at any band.} \tablefoottext{b}{Variability parameter obtained from Eq. \ref{eq_alpha_var}.} \tablefoottext{c}{Colliding-wind binary candidates: p = primary candidate, s = secondary candidate}}

\end{table*}

\begin{figure}
  \resizebox{\hsize}{!}{\includegraphics{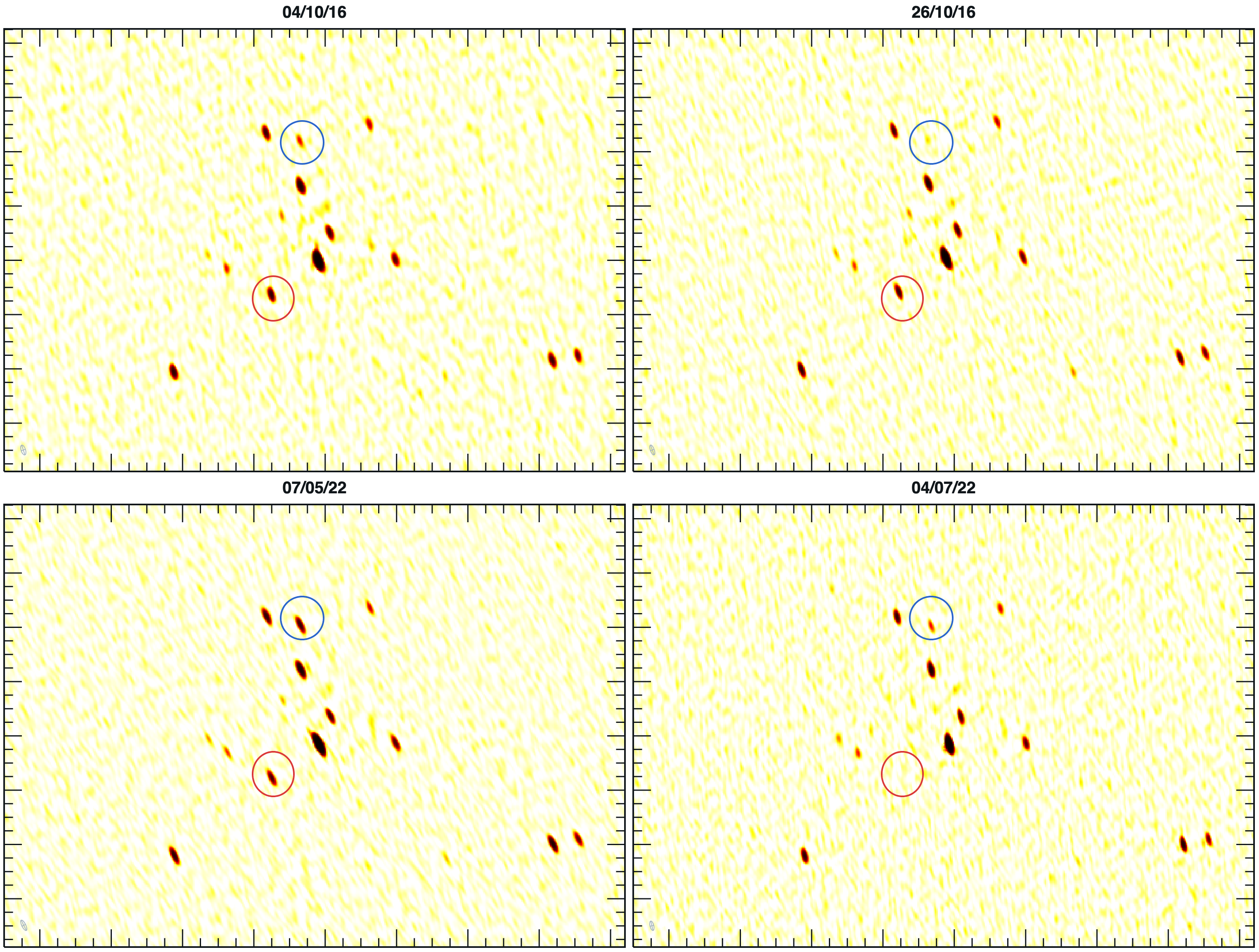}}
  \caption{Close up of the Arches cluster X-band images from 2016 (top) and 2022 (bottom). The blue and red circles shows the position of F18 and F26 respectively.}
  \label{fig_F18_F26}
\end{figure}

In the light of these results, we can confidently categorise F6, F19, F18 and F26 as primary CWB candidates due to their consistent $\alpha\lesssim 0.0$ values across the time span of our observations. Furthermore, F18 and F26 are clear examples of extreme variability (see Fig \ref{fig_F18_F26}) within months which further strengthens their binary status. In fact, F26 flux falls below the rms detection threshold ($5.4\, \mu$Jy/beam) for the 4 July observations in X-band. Such a drastic change in observed flux is not expected from the emission arising from the stellar winds of isolated stars.

As previously mentioned, some sources with $\alpha>0$ can still be considered secondary binary candidates. By looking at strong changes in flux $(\sigma_{\mathrm{VAR}}\gtrsim10)$ over different observing epochs, we may categorise F7, and F9 as secondary CWB candidates. This hypothesis is further supported by the fact that these sources show clear X-ray counterparts detected by \textit{Chandra} \citep{Muno2009}. Moreover, F7 has been identified as a long period ($P\gtrsim1200$ days) radial velocity binary candidate by \citet{Clark_IV}. Table \ref{table_Chandra} shows the properties of the \textit{Chandra} counterparts to the radio detections.

\begin{table}
\caption{Hardness ratios of the sources detected by \textit{Chandra}.}
\label{table_Chandra}
\centering
\begin{tabular}{c c c c}
\hline \hline
ID  & HR0 &  HR1 & HR2 \\
\hline
     F6 & $0.769_{-0.056}^{+0.053}$ & $0.023_{-0.063}^{+0.064}$ & $-0.133_{-0.066}^{+0.065}$ \\
     F7 & $0.760_{-0.066}^{+0.064}$ & $-0.042_{-0.077}^{+0.078}$ & $-0.270\pm0.086$ \\
    F2/F19 & $>0.336$ & $0.564_{-0.358}^{+0.390}$ & $0.114_{-0.313}^{+0.316}$ \\
     F9 & $0.773\pm0.069$ & $0.063_{-0.079}^{+0.080}$ & $-0.016\pm0.078$ \\
\hline \hline
\end{tabular}
\tablefoot{See \citet{Muno2009} for the definition of HR0, HR1 and HR2.}
\end{table}

It is possible that isolated massive stars produce thermal X-ray emission. Such emission can be caused by a difference in the velocity of two wind shocks travelling radially outwards from the stellar surface. However, the X-ray flux associated with such emission is estimated to be more than an order of magnitude lower than the one produced in the colliding wind region of a binary system \citep{DeBecker2007} and therefore not in the detection limit of current X-ray observatories such as \textit{Chandra} at the distance of the GC.

Finally, F12, F14, and F55 also show negative spectral indices (or compatibility with a flat spectrum within the uncertainties) at some point during our observations. In particular, F14 seems to show a change in the main emission mechanism from non-thermal during 2018 to thermal during 2022 observations. This could be indicating a binary system with a long orbital period, with the colliding wind region changing from optically thin to optically thick. However, due to their relatively faint flux, specially in C-band which is more prone to contamination from extended emission, we categorise these sources as secondary CWB candidates. Further monitoring of the Arches members with high-sensitivity, high-resolution radio-interferometers such as the VLA or the future SKA-MID may fully uncover the nature of the emission associated with these radio-stars.

In all, out of the 23 radio-stars we detect, we infer a binarity fraction of $9/23\approx39\%$ with radio data alone, which increases to $14/23\approx61\%$ if we take into account the radial velocity candidates from \citet{Clark_IV}. Furthermore, when we compare our candidates to theirs, we can see that, with radio observations, we are able to discriminate two binary candidates F19 and F26, that did not meet the significance threshold of radial velocity difference in their work, as well as add another candidate F55, not present in their analysis.

\subsubsection{Thermal radio-stars}
We can clearly categorise F3, Dong19, F5, F8, Dong96, and B1 as thermal sources whose emission is dominated by free-free emission from their stellar winds based on their consistently positive spectral indices and relatively low variability. However, it is still possible that the strong, optically thick winds of these Wolf-Rayet stars eclipse a potential non-thermal contribution from the colliding wind region of a multiple system, specially in short-period binaries \citep{DeBecker2007}.

\subsubsection{Ambiguous cases and other radio sources}
The radio-stars F1 and F2, show thermal spectral indices for most of the observations. However, in the case of F1, the X-band observations carried out on 11 April 2018 return a negative spectral index value (albeit with considerable uncertainties) and the 4 July 2022 observations show an spectral index considerably lower than the canonically thermal value of $\alpha\approx0.6$, which may indicate a possible non-thermal component. Similar results are obtained for F2. The fact that this is a short period ($P\sim10$ day) binary with a low eccentricity $(e=0.075)$ orbit \citep{Lohr2018} may explain why we do not detect clear traces of synchrotron emission in our observations with the exception of those carried out during 7 May 2022, whose spectral index is compatible with a flat spectrum, although uncertainties are large. Therefore, we cannot confidently classify these sources as CWB candidates with these results. In order to more reliably identify the nature of these radio-stars, more 'simultaneous' multi-band observations are needed, resulting in more accurate computations of their spectral indices (note that the spectral indices derived from the 4 July 2022 C- and X-band combined data have the best overall uncertainties in Table \ref{table_alphas}). Finally, despite showing significant variability, we categorise F4 as a thermal emitter, given that its spectral index is consistently thermal in all our observations. A plausible explanation for F4 variability could be a big over-density in the winds in the radio emitting region. We deem unlikely that F4 underwent an eruptive mass-loss episode given its current evolutionary status, as these eruptive events are expected from luminous blue variables \citep[e.g.~][]{Jiang2018}, but not in WNh stars. 

The rest of the detections not mentioned in the above sub-sections are too faint to be subject to the least squares fitting from Sect. \ref{subsection_alphas} or not detected in a particular band. Namely: F16, F10, F15, F17, AR19, F28, and AR20. Most of them are $5\sigma$ detections in combined images or in the deep images. Given that they are mostly detected in X-band, we assumed thermal emission and included most of them in the mass-loss analysis (see Table \ref{table_MLRs}). Furthermore, AR19 and AR20 are only detected in the X-band deep image, and they present no clear NIR stellar counterpart. For these reasons, they are excluded from the analysis presented in this work. In addition, some slightly extended sources near the cluster core can be seen in Fig. \ref{Arches_deep_X}. As explained in Sect. \ref{section_obs}, such sources are most likely remnants of the extended emission that were not fully removed by the \textit{u-v} cut, as they do not present NIR stellar counterparts in \citet{Hosek2022} catalogue, and do not match the synthesised beam shape in any image. More observations carried out at different array configurations may better characterise such emission.

In the future, we will explore the full field of view of the observations and the properties of the non-stellar sources lying within the VLA primary beam.

\section{Summary and conclusions}\label{sect_conclusions}
We present the most complete radio-stellar catalogue (total of 23 radio-stars) and the deepest (rms noise $\sim2.5\, \mu\mathrm{Jy/beam}$ in the case of X-band deep image) radio continuum C- and X-band images of the Arches cluster to date. With observations scattered across six years, the VLA sensitivity and angular resolution have considerably improved our understanding of radio-stars within the Arches cluster.

We find that around $60\%$ of the radio-stars in the Arches show significant flux variability on timescales of years. Furthermore, non-thermal sources are more likely to show drastic changes in flux densities in timescales of months. 

We have also derived the clumping-scaled mass-loss rates of our thermal sources. They show values consistent with the existing literature. In addition, we provide a list of clumping ratios for our thermal sources. In general, according to the clumping ratios, C- and X-band are similarly affected by clumping (ratios close to unity), but it may become more prominent at higher frequencies.

We classify four of our sources (F6, F18, F19, and F26) as CWBs, as they display $\alpha\lesssim0.0$ across our observations. In addition, we show preliminary evidence that high radio-flux or/and spectral index variability may be an indirect indicator of multiplicity, an hypothesis that is strengthened by multi-wavelength counterparts to our radio detections (NIR radial velocity studies and X-ray point sources). Therefore, we categorise F7, F9, F12, F14, and F55 as secondary CWB candidates, expanding the number of CWB candidates by four (F26 as primary candidate and F7, F14, and F55 as secondary candidates) when compared to our pilot study \citep{G-C2021}. Most of the remaining radio-stars for which spectral indices could be derived (12/18) show $\alpha\sim0.6$, consistent with thermal emission being the dominant mechanism at work.  Thus, combining our CWB candidates with those from the radial velocity study of \citet{Clark_IV}, we find that $14/23\approx61\%$ of the radio-stars of the Arches cluster are binary (or higher order multiples) candidates.

We show preliminary evidence of spectral index variation in both thermal and non-thermal radio-stars. Significant changes in $\alpha$ in the case of thermal emitters may be an indicator of changes in the clumping structure of their stellar winds or (more unlikely) eruptive episodes of mass-loss. Observed changes of flux and/or spectral index in non-thermal sources may be related to the orbital phases of binaries, but currently the data are still too incomplete for a conclusive study. Therefore, we emphasise the need for more surveys with more frequent, long term sampling and simultaneous multi-band observations to disentangle the thermal and non-thermal contribution to the emission of a CWB candidate, and thus help constraining the multiplicity of massive stars in regions of the parameter space not available for radial velocity studies. 

Moreover, more simultaneous multi-wavelength observations may help sampling wind clumping at different atmospheric heights. Near infrared, millimetre and sub-millimetre radio observations, and centimetre radio continuum studies could provide unprecedented detail into the clumping of massive stellar winds and thus improve our understanding of mass-loss and its effect on the poorly understood evolution of post main-sequence massive stars.

This study shows the viability of radio observations to discern thermal and non-thermal emission mechanisms of radio-stars at the GC. Further monitoring covering a dense temporal baseline could help unravel the relationship between the dominant emission mechanism and the orbital phase of a multiple system, establishing radio continuum observations as a reliable and complementary method to study the multiplicity fraction of young massive clusters.

\begin{acknowledgements} 
We thank the NRAO staff for their help setting up the observations and their excellent guidance with the VLA pipeline and calibration process. Authors MCG, RS, AA, JM, MPT, and ATGC acknowledge financial support from the Severo Ochoa grant CEX2021-001131-S funded by MCIN/AEI/ 10.13039/501100011033. MGC and RS acknowledge support from grant EUR2022-134031 funded by MCIN/AEI/10.13039/501100011033 and by the European Union NextGenerationEU/PRTR. and by grant PID2022- 136640NB-C21 funded by MCIN/AEI 10.13039/501100011033 and by the European Union.

AA and MPT acknowledge support from the Spanish National grant 
PID2020-117404GB-C21, funded by MCIN/AEI/10.13039/501100011033.

ATGC acknowledges the Astrophysics and High Energy Physics programme supported by MCIN with funding from European Union NextGenerationEU (PRTR-C17.I1) and by Generalitat Valenciana.

F.N., acknowledges support by grants PID2019-105552RB-C41 and PID2022-137779OB-C41 funded by
MCIN/AEI/10.13039/501100011033 by "ERDF A way of making Europe".

Authors MCG, RS and JM acknowledge the Spanish Prototype of an SRC (SPSRC) service and support funded by the Ministerio de Ciencia, Innovación y Universidades (MICIU), by the Junta de Andalucía, by the European Regional Development Funds (ERDF) and by the European Union NextGenerationEU/PRTR. The SPSRC acknowledges financial support from the Agencia Estatal de Investigación (AEI) through the "Center of Excellence Severo Ochoa" award to the Instituto de Astrofísica de Andalucía (IAA-CSIC) (SEV-2017-0709) and from the grant CEX2021-001131-S funded by MICIU/AEI/ 10.13039/501100011033.

JM acknowledges financial support from the grant  PID2021-123930OB-C21 funded by MICIU/AEI/ 10.13039/501100011033 and by ERDF/EU.

\end{acknowledgements}

\bibliographystyle{aa} 
\bibliography{references.bib} 

\begin{appendix}
\section{Additional material}
\def\arraystretch{1.2}
\begin{table*}
\caption{Observations and image properties.}
\label{table_observations}      
\centering
\begin{tabular}{c c c c c}
\hline \hline
Date  &   Band & ($\theta_{\mathrm{maj}}^\mathrm{FWHM}\times\theta_{\mathrm{min}}^{\mathrm{FWHM}},\ \mathrm{PA}$)  & Off-source rms ($\mu$Jy/beam)& Time on target (minutes)\\    
\hline                        
4 Oct 2016   & X & $0\farcs38\times0\farcs17,\ 18\degr$ & $5.0$ &  52\\      
26 Oct 2016   & X & $0\farcs39\times0\farcs14,\ 20\degr$  & $5.2$ & 47\\
combined 2016  & X & $0\farcs38\times0\farcs15,\ 19\degr$ & $3.7$ & 99\\
11 Apr 2018  & X & $0\farcs39\times0\farcs14,\ -16\degr$ & $4.6$ &  47\\      
10 Jun 2018  & C & $0\farcs51\times0\farcs23,\ 0\degr$  & $5.6$ & 76\\
7 May 2022   & X & $0\farcs42\times0\farcs15,\ 26\degr$ & $4.1$ &  86\\      
4 Jul 2022  & X & $0\farcs36\times0\farcs14,\ 14\degr$   & $5.4$ & 62\\
combined 2022   & X & $0\farcs38\times0\farcs14,\ 20\degr$     & $3.2$ & 148\\
4 Jul 2022  & C & $0\farcs55\times0\farcs23,\ -6\degr$    & $5.9$ & 62\\
\hline
Deep X-band   & X & $0\farcs31\times0\farcs14,\ 11\degr$     & $2.5$ & 294\\
Deep C-band   & C & $0\farcs52\times0\farcs23,\ -3\degr$     & $4.4$ & 138\\
\hline \hline

\end{tabular}
\tablefoot{All data were acquired in A-configuration.}
\end{table*}



\setlength{\tabcolsep}{4pt}
\begin{sidewaystable*}
\footnotesize
\begin{center}
\caption{Arches radio sources}
\label{fluxtable}
\resizebox{1.0\textwidth}{!}{%
\begin{tabular}{ccccccccccccccccc}
\hline  \hline
\noalign{\vskip 1mm}
ID\tablefootmark{(a)} & Type\tablefootmark{(b)}   &  R.A. ($\degr$) & eR.A. ($\arcsec$) & $\delta$ ($\degr$) &  e$\delta$ ($\arcsec$) & $S_X^{\mathrm{deep}}$ &  $S_C^{\mathrm{deep}}$ & $S_X^{04/10/16}$ & $S_X^{26/10/16}$ &  $S_X^{2016}$ & $S_X^{11/04/18}$ & $S_X^{07/05/22}$ & $S_X^{04/07/22}$ & $S_X^{2022}$ & $S_C^{10/06/18}$ & $S_C^{04/07/22}$ \\ 
\noalign{\vskip 0.5mm}

\hline
     F6 & WN8-9h + ? & 266.460064 & 0.004 & -28.822858 & 0.013 & 2.277 $\pm$ 0.069 & 2.292 $\pm$ 0.117 & 2.066 $\pm$ 0.064 & 2.050 $\pm$ 0.063 & 2.069 $\pm$ 0.068 & 2.170 $\pm$ 0.066 & 2.393 $\pm$ 0.074 & 2.281 $\pm$ 0.070 & 2.361 $\pm$ 0.071 & 2.171 $\pm$ 0.110 & 2.492 $\pm$ 0.127 \\
     F7 & WN8-9h + ? & 266.460275 & 0.004 & -28.822092 & 0.013 & 0.422 $\pm$ 0.015 & 0.258 $\pm$ 0.032 & 0.373 $\pm$ 0.017 & 0.350 $\pm$ 0.014 & 0.369 $\pm$ 0.015 & 0.363 $\pm$ 0.015 & 0.445 $\pm$ 0.020 & 0.417 $\pm$ 0.018 & 0.440 $\pm$ 0.016 & 0.242 $\pm$ 0.028 & 0.260 $\pm$ 0.025 \\
    F19 & O4-5 Ia & 266.457327 & 0.005 & -28.823879 & 0.013 & 0.291 $\pm$ 0.010 & 0.396 $\pm$ 0.030 & 0.211 $\pm$ 0.011 & 0.204 $\pm$ 0.011 & 0.210 $\pm$ 0.009 & 0.285 $\pm$ 0.012 & 0.321 $\pm$ 0.014 & 0.306 $\pm$ 0.015 & 0.316 $\pm$ 0.011 & 0.419 $\pm$ 0.023 & 0.365 $\pm$ 0.021 \\
     F3 & WN8-9h & 266.461753 & 0.004 & -28.823997 & 0.013 & 0.264 $\pm$ 0.009 & 0.181 $\pm$ 0.013 & 0.258 $\pm$ 0.012 & 0.246 $\pm$ 0.011 & 0.258 $\pm$ 0.011 & 0.252 $\pm$ 0.012 & 0.272 $\pm$ 0.011 & 0.264 $\pm$ 0.014 & 0.267 $\pm$ 0.010 & 0.189 $\pm$ 0.014 & 0.172 $\pm$ 0.014 \\
 Dong19 & WN8-9h & 266.452459 & 0.008 & -28.828502 & 0.015 & 0.239 $\pm$ 0.008 & 0.157 $\pm$ 0.011 & 0.231 $\pm$ 0.011 & 0.218 $\pm$ 0.011 & 0.229 $\pm$ 0.010 & 0.221 $\pm$ 0.009 & 0.237 $\pm$ 0.010 & 0.239 $\pm$ 0.011 & 0.239 $\pm$ 0.009 & 0.156 $\pm$ 0.013 & 0.159 $\pm$ 0.013 \\
     F4 & WN7-8h & 266.460674 & 0.004 & -28.821549 & 0.013 & 0.237 $\pm$ 0.009 & 0.147 $\pm$ 0.011 & 0.204 $\pm$ 0.010 & 0.198 $\pm$ 0.010 & 0.208 $\pm$ 0.009 & 0.214 $\pm$ 0.010 & 0.255 $\pm$ 0.011 & 0.246 $\pm$ 0.013 & 0.250 $\pm$ 0.009 & 0.127 $\pm$ 0.010 & 0.162 $\pm$ 0.013 \\
     F5 & WN8-9h & 266.460199 & 0.005 & -28.825529 & 0.013 & 0.227 $\pm$ 0.008 & 0.150 $\pm$ 0.010 & 0.205 $\pm$ 0.010 & 0.212 $\pm$ 0.010 & 0.213 $\pm$ 0.009 & 0.209 $\pm$ 0.010 & 0.226 $\pm$ 0.011 & 0.225 $\pm$ 0.013 & 0.226 $\pm$ 0.009 & 0.154 $\pm$ 0.011 & 0.146 $\pm$ 0.012 \\
     F8 & WN8-9h & 266.459930 & 0.004 & -28.822570 & 0.014 & 0.216 $\pm$ 0.014 & 0.127 $\pm$ 0.028 & 0.217 $\pm$ 0.025 & 0.216 $\pm$ 0.036 & 0.223 $\pm$ 0.026 & 0.204 $\pm$ 0.025 & 0.207 $\pm$ 0.027 & 0.214 $\pm$ 0.040 & 0.212 $\pm$ 0.018 & 0.163 $\pm$ 0.028 & 0.152 $\pm$ 0.031 \\
     Dong96 & WN8 & 266.452311 & 0.01 & -28.835008 & 0.01 & 0.187 $\pm$ 0.009 & 0.126 $\pm$ 0.012 & $0.175 \pm 0.011$ & $0.191 \pm 0.011$ & $0.183 \pm 0.009$ & $0.182 \pm 0.009$ & $0.193 \pm 0.010$ & $0.186 \pm 0.012$ & $0.192 \pm 0.009$ & $0.126 \pm 0.012$ & $0.124 \pm 0.012$ \\ 
     F1 & WN8-9h & 266.459165 & 0.004 & -28.822847 & 0.013 & 0.180 $\pm$ 0.008 & 0.106 $\pm$ 0.021 & 0.165 $\pm$ 0.011 & 0.163 $\pm$ 0.011 & 0.166 $\pm$ 0.010 & 0.175 $\pm$ 0.011 & 0.184 $\pm$ 0.011 & 0.177 $\pm$ 0.013 & 0.184 $\pm$ 0.009 & 0.104 $\pm$ 0.016 & 0.110 $\pm$ 0.019 \\
    F18 & O4-5 Ia$^{+}$ & 266.460279 & 0.004 & -28.821638 & 0.014 & 0.157 $\pm$ 0.007 & 0.168 $\pm$ 0.016 & 0.070 $\pm$ 0.010 & 0.028 $\pm$ 0.009 & 0.053 $\pm$ 0.008 & 0.198 $\pm$ 0.011 & 0.229 $\pm$ 0.011 & 0.068 $\pm$ 0.010 & 0.180 $\pm$ 0.009 & 0.252 $\pm$ 0.017 & 0.080 $\pm$ 0.016 \\
    F26 & O4-5 Ia & 266.460613 & 0.004 & -28.823205 & 0.014 & 0.152 $\pm$ 0.008 & -- & 0.218 $\pm$ 0.011 & 0.209 $\pm$ 0.011 & 0.222 $\pm$ 0.010 & 0.174 $\pm$ 0.010 & 0.184 $\pm$ 0.011 & -- & 0.124 $\pm$ 0.009 & -- & -- \\
     F2 & WN8-9h + O5-6 Ia$^{+}$ & 266.457032 & 0.005 & -28.823828 & 0.014 & 0.146 $\pm$ 0.006 & 0.100 $\pm$ 0.014 & 0.144 $\pm$ 0.010 & 0.143 $\pm$ 0.010 & 0.148 $\pm$ 0.008 & 0.152 $\pm$ 0.010 & 0.146 $\pm$ 0.010 & 0.132 $\pm$ 0.011 & 0.142 $\pm$ 0.007 & 0.113 $\pm$ 0.011 & 0.105 $\pm$ 0.016 \\
     F9 & WN8-9h & 266.459402 & 0.004 & -28.819929 & 0.014 & 0.139 $\pm$ 0.006 & 0.059 $\pm$ 0.009 & 0.237 $\pm$ 0.011 & 0.241 $\pm$ 0.012 & 0.239 $\pm$ 0.009 & 0.120 $\pm$ 0.010 & 0.114 $\pm$ 0.008 & 0.093 $\pm$ 0.011 & 0.109 $\pm$ 0.007 & 0.066 $\pm$ 0.011 & 0.055 $\pm$ 0.010 \\
    F12 & WN7-8h & 266.459466 & 0.004 & -28.821461 & 0.014 & 0.086 $\pm$ 0.006 & 0.060 $\pm$ 0.009 & 0.094 $\pm$ 0.012 & 0.092 $\pm$ 0.009 & 0.098 $\pm$ 0.008 & 0.069 $\pm$ 0.007 & 0.082 $\pm$ 0.008 & 0.085 $\pm$ 0.010 & 0.083 $\pm$ 0.007 & 0.051 $\pm$ 0.008 & 0.072 $\pm$ 0.011 \\
     B1 & WN8-9h & 266.464398 & 0.005 & -28.823810 & 0.015 & 0.069 $\pm$ 0.005 & 0.049 $\pm$ 0.009 & 0.056 $\pm$ 0.008 & 0.070 $\pm$ 0.008 & 0.063 $\pm$ 0.006 & 0.072 $\pm$ 0.008 & 0.067 $\pm$ 0.007 & 0.063 $\pm$ 0.009 & 0.067 $\pm$ 0.006 & 0.045 $\pm$ 0.010 & 0.044 $\pm$ 0.010 \\
    F14 & WN8-9h & 266.461134 & 0.007 & -28.822939 & 0.020 & 0.058 $\pm$ 0.009 & 0.059 $\pm$ 0.012 & 0.066 $\pm$ 0.011 & 0.058 $\pm$ 0.011 & 0.063 $\pm$ 0.010 & 0.055 $\pm$ 0.011 & 0.053 $\pm$ 0.010 & 0.063 $\pm$ 0.012 & 0.043 $\pm$ 0.013 & 0.124 $\pm$ 0.017 & 0.044 $\pm$ 0.011 \\
    F55 & O5.5-6 III & 266.461351 & 0.012 & -28.822806 & 0.033 & 0.038 $\pm$ 0.011 & 0.052 $\pm$ 0.013 & 0.024 $\pm$ 0.013 & 0.032 $\pm$ 0.012 & 0.028 $\pm$ 0.012 & 0.030 $\pm$ 0.011 & 0.041 $\pm$ 0.010 & 0.049 $\pm$ 0.015 & 0.041 $\pm$ 0.012 & 0.054 $\pm$ 0.015 & 0.076 $\pm$ 0.021 \\
    F16 & WN8-9h & 266.460493 & 0.007 & -28.822400 & 0.024 & 0.030 $\pm$ 0.006 & -- & 0.038 $\pm$ 0.009 & 0.037 $\pm$ 0.009 & 0.039 $\pm$ 0.008 & 0.032 $\pm$ 0.010 & 0.029 $\pm$ 0.007 & -- & 0.025 $\pm$ 0.006 & -- & -- \\
    F10 & O7-8 Ia$^{+}$ & 266.458574 & 0.006 & -28.824029 & 0.019 & 0.029 $\pm$ 0.004 & -- & 0.024 $\pm$ 0.008 & 0.042 $\pm$ 0.009 & 0.029 $\pm$ 0.006 & 0.046 $\pm$ 0.010 & 0.036 $\pm$ 0.007 & -- & 0.030 $\pm$ 0.006 & -- & -- \\
    F15 & O6-7 Ia$^{+}$ + ? & 266.461446 & 0.006 & -28.821275 & 0.021 & 0.017 $\pm$ 0.004 & -- & -- & -- & 0.023 $\pm$ 0.007 & -- & 0.020 $\pm$ 0.005 & -- & 0.022 $\pm$ 0.005 & -- & -- \\
    F17 & O5-6 Ia$^{+}$ & 266.458872 & 0.013 & -28.824218 & 0.024 & 0.021 $\pm$ 0.005 & -- & -- & -- & 0.016 $\pm$ 0.005 & -- & -- & -- & 0.036 $\pm$ 0.009 & -- & -- \\
   AR19 & -- & 266.462521 & 0.008 & -28.820484 & 0.024 & 0.015 $\pm$ 0.004 & -- & -- & -- & -- & -- & -- & -- & -- & -- & -- \\
    F28 & O4-5 Ia & 266.460980 & 0.005 & -28.822704 & 0.024 & 0.013 $\pm$ 0.003 & -- & -- & -- & -- & -- & -- & -- & 0.016 $\pm$ 0.005 & -- & -- \\
   AR20 & -- & 266.460587 & 0.030 & -28.823940 & 0.042 & 0.023 $\pm$ 0.008 & -- & -- & -- & -- & -- & -- & -- & -- & -- & -- \\
\hline \hline

\end{tabular}
}
\tablefoot{\footnotesize{All flux densities and their related uncertainties are in mJy units. Date format: $S_\nu^{\mathrm{[DD/MM/YY]}}$ for individual observations, $S_\nu^{\mathrm{[YEAR]}}$ for combined images.
   \tablefoottext{a}{Stellar ID from \citet{Clark_I} or \citet{Dong2011}.}
   \tablefoottext{b}{Spectral type from \citet{Clark_I,Clark_IV} or \citet{Dong2011}.}}}
\end{center}
\end{sidewaystable*}

\FloatBarrier


\begin{figure*}
\centering
  \includegraphics[width=17cm]{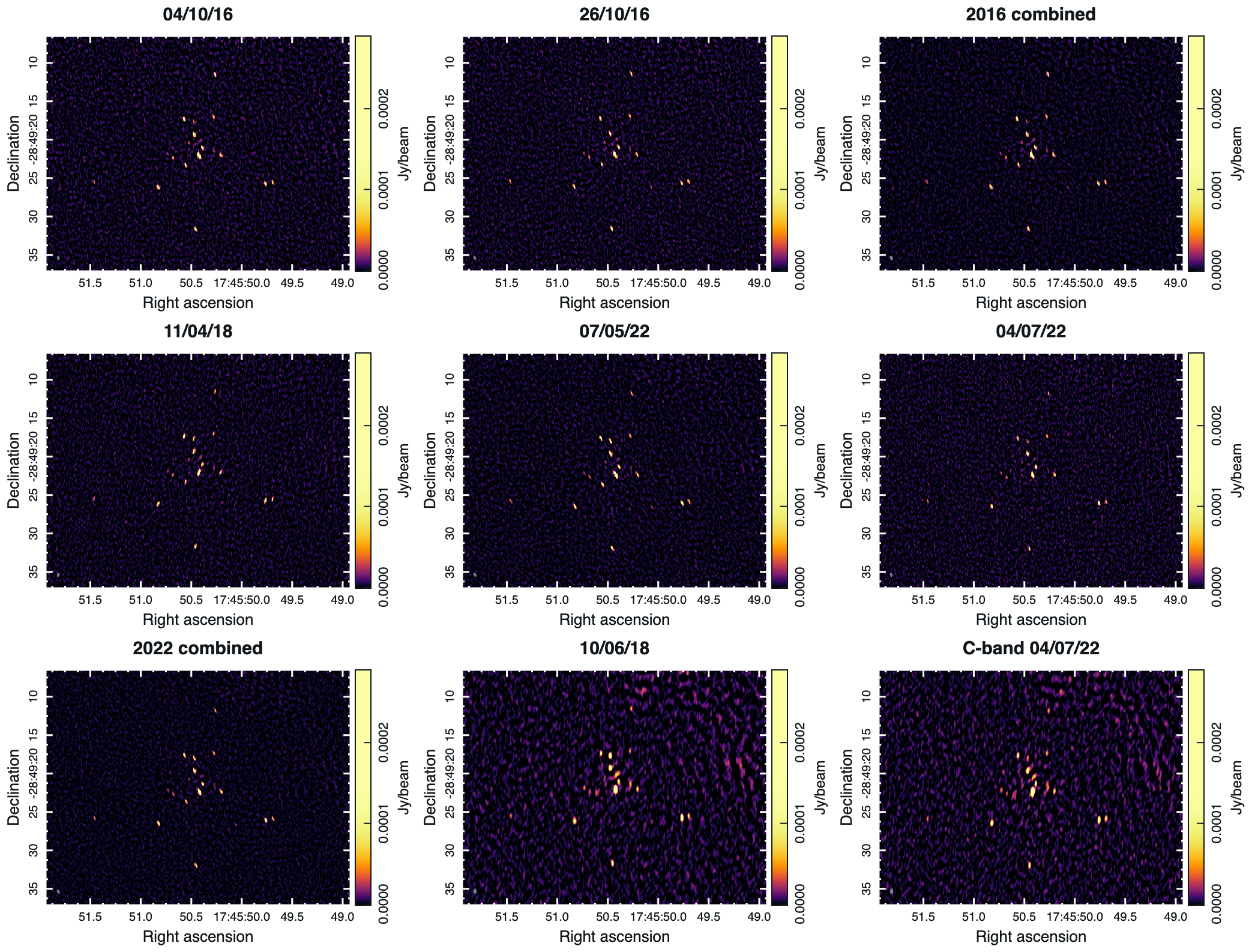}
    \caption{All images of the Arches cluster.}
    \label{<Your label>}
\end{figure*}
\end{appendix}

\end{document}